\documentclass[aps,prl,superscriptaddress,floatfix,10pt,twocolumn]{revtex4-1}
\usepackage{graphicx,graphics}
\usepackage{dcolumn}
\usepackage{amsmath,amssymb,amsfonts}
\usepackage{latexsym,verbatim}
\usepackage{bm,dsfont}
\usepackage{color}
\usepackage{xcolor}
\usepackage{ulem}
\usepackage{multirow}
\usepackage[breaklinks=true,colorlinks,citecolor=blue,linkcolor=blue,urlcolor=blue]{hyperref}
\usepackage{array}
\usepackage{nicefrac}
\usepackage{overpic}

\begin{document}
\title{Globally driven superconducting quantum computing architecture}
\author{Roberto Menta}
\email{rmenta@planckian.co}
\thanks{These authors contributed equally to this work.}
\affiliation{Planckian srl, I-56127 Pisa, Italy}
\affiliation{NEST, Scuola Normale Superiore, I-56127 Pisa, Italy}
\author{Francesco Cioni}
\email{francesco.cioni@sns.it}
\thanks{These authors contributed equally to this work.}
\affiliation{NEST, Scuola Normale Superiore, I-56127 Pisa, Italy}
\author{Riccardo Aiudi}
\affiliation{Planckian srl, I-56127 Pisa, Italy}
\author{Marco Polini}
\affiliation{Planckian srl, I-56127 Pisa, Italy}
\affiliation{Dipartimento di Fisica dell’Universit\`{a} di Pisa, Largo Bruno Pontecorvo 3, I-56127 Pisa, Italy}
\author{Vittorio Giovannetti}
\affiliation{Planckian srl, I-56127 Pisa, Italy}
\affiliation{NEST, Scuola Normale Superiore, I-56127 Pisa, Italy}

\begin{abstract}
We propose a platform for implementing a universal, {\it globally driven} quantum computer based on a 2D ladder hosting three different species of superconducting qubits. In stark contrast with the existing literature, our scheme exploits the always-on longitudinal ZZ coupling. The latter, combined with specific driving frequencies, enables the reach of a blockade regime, which plays a pivotal role in the computing scheme.
\end{abstract}

\maketitle

\paragraph{Introduction.}
Superconducting circuits have demonstrated exceptional performance in executing precise control and measurement operations, thus making them a preferred choice for quantum computing (QC) architectures~\cite{Gambetta2017,Martinis2019quantum,SQ2020, supremacy_PRL_2021, blais_review_RMP,bravyi2022future,ezratty2023perspective}.
Nevertheless, notwithstanding their remarkable properties, these platforms are still afflicted by a serious drawback.
It is indeed widely recognized that the scalability of superconducting (and other solid-state) architectures faces a hurdle because of the need for {\it localized} control of each logical qubit~\cite{You2005supercond, Girvin2008, devoret2013supercond, wedin2017quantum}. This challenge, often dubbed the ``wiring problem", arises from the necessity of multiple control signals for each qubit in standard QC architectures, leading to a wiring overload~\cite{Gambetta2017, tsai2020pseudo, tsai2021gate}. In view of this fact, maintaining high gate fidelity while scaling up the number of qubits within a single processor presents a significant hurdle~\cite{Zhang2022}. 
While state-of-the-art superconducting QC platforms can achieve single-qubit operation fidelities as high as $99.99 \%$~\cite{Zhiyuan2023,Somoroff2023}, reducing errors in two-qubit gates remains challenging. To our knowledge, the associated error rates for these operations persist at around $0.1\%$~\cite{Barends2019, GoogleAIQuantum,Wallraff2020,Ding2023,Singh2024}. Incidentally, one of  the main limiting factors to enhance  two-qubit gate fidelity
in superconducting  platforms  is the ``residual'' longitudinal ZZ interaction between neighboring qubits. While recent research endeavors have showcased methods to alleviate~\cite{ZZ-suppression2020,ZZ-eliminating2022,xu2021z,Huang2023} and even leverage ZZ coupling~\cite{ZZ-contrast2020,ZZ_2020, ZZ_2021} for implementing two-qubit gates,
such interaction remains generally undesirable within conventional quantum superconducting QC frameworks.
\begin{figure}[t!]
\centering
\includegraphics[width=1.0\columnwidth]{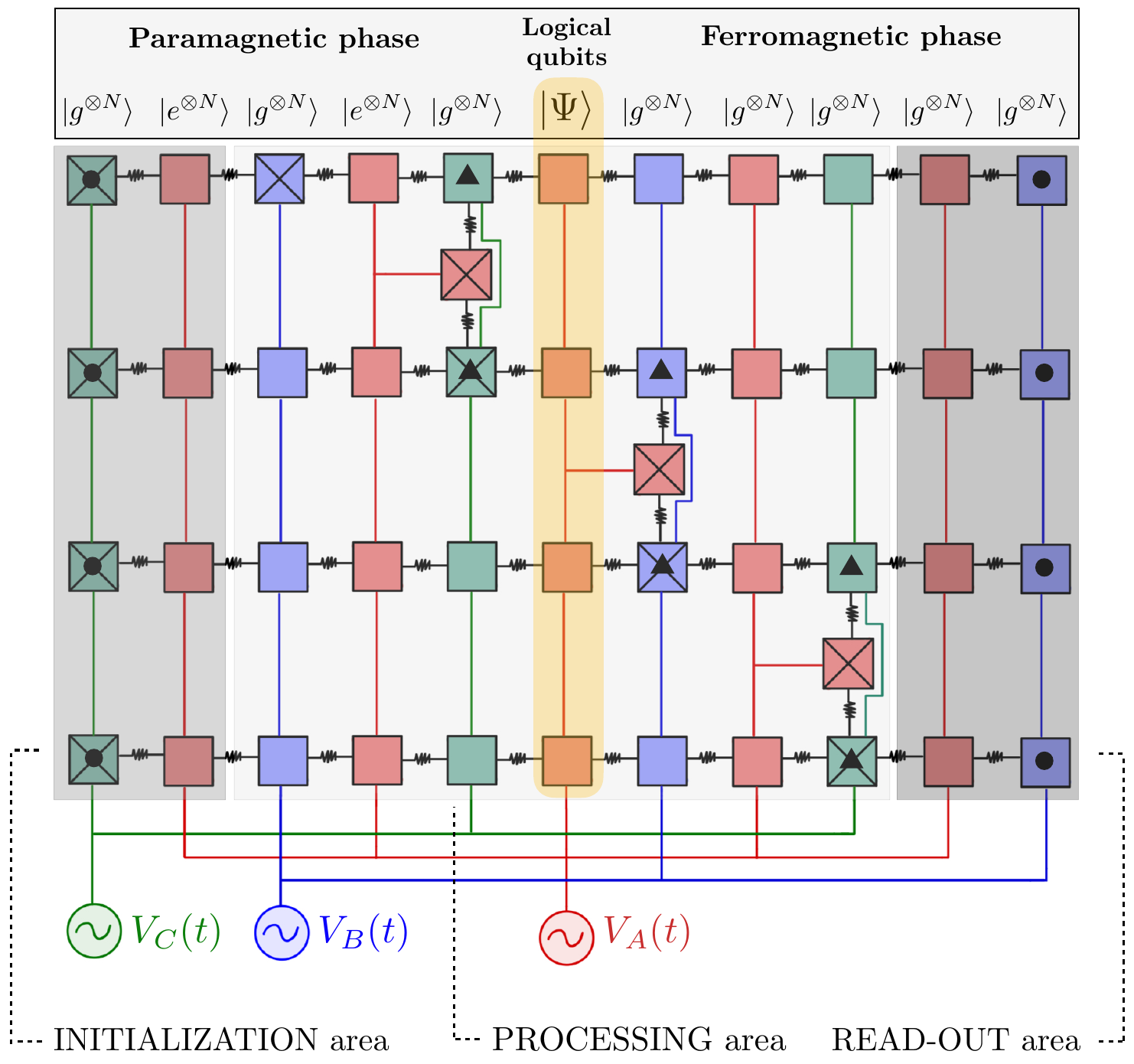}
\caption{Architecture for a globally driven superconducting quantum computer. 
Three groups, $A$, $B$, $C$ of superconducting qubits (red, blue, and green squares, respectively) occupy the horizontal rows of a 2D ladder. Black springs, colored continuous lines, crosses, filled circles, and triangles are defined in the main text. Electrical wires depicted as continuous red, blue, and green lines connect all elements within each group, facilitating global control through independent classical electrical pulses. During the computation, the logical information is encoded in the qubits of one of the columns of the processing area (enlightened in yellow in the figure): qubits on the left (right)-hand side of such information carrier column  are in a ``paramagnetic'' (``ferromagnetic'') $geg$ ($ggg$) phase. The illustration pertains to a $N=4$ logical qubit  quantum computer.
\label{fig:transarray}}
\end{figure}

To offer a solution to these daunting issues,  we present a universal superconducting QC architecture where, exploiting the presence of always-on ZZ coupling terms, local driving of each individual qubit is replaced by {\it global} pulses that control {\it collectively} a large collection (roughly one third) of the memory registers of the model, hence drastically reducing the total number of wires necessary to run the computation. Historically, a pioneering global scheme of QC was first proposed by Lloyd~\cite{Lloyd_1993}. Soon after, several other proposals for QC based on global control were put forward~\cite{benjamin_2000, benjamin_2001, benjamin_2003}. However, these failed to reach the readiness level necessary to be competitive with models based on local control. Our scheme is inspired by a recent disruptive advancement that was  made by Cesa and Pichler~\cite{cesa2023universal}, who presented a universal quantum computer  based on globally driven Rydberg atoms. 
In this QC platform~\cite{Saffman_RMP_2010,fromonteil2024hamilton}, two nearby atoms in an excited Rydberg state interact so strongly that their simultaneous excitation is forbidden (``Rydberg blockade''~\cite{Rydberg_blockade}).
Building on this effect, the authors of Ref.~\cite{cesa2023universal}  demonstrated that by arranging {\it two} different species of ${\cal O}(N^2)$ Rydberg atoms in an almost-regular  rectangular lattice,  one can implement globally driven QC on $N$ qubits.

Here, we propose a generalization of the results of Ref.~\cite{cesa2023universal} to a solid-state, superconducting platform. Readers will immediately understand that this is far from trivial since the aforementioned blockade effect does not come ``for free'' in our platform. Below, we show how to {\it dynamically} induce blockade into a lattice of superconducting qubits. Additionally, we use {\it three} rather than {\it two} species of qubits, which, as we argue below, significantly reduces (with respect to Ref.~\cite{cesa2023universal}) the number of physical qubits that are needed to implement a universal quantum computer. Last but not least, we propose a different initialization scheme with respect to that of Ref.~\cite{cesa2023universal}, which helps us reducing the size of the initialization area of our solid-state quantum computer with respect to the Rydberg one. In our model, quantum information is always localized in the $N$ superconducting qubits of one of the vertical columns that form the lattice (see Fig.~\ref{fig:transarray}), referred to as the ``information carrier column'' (ICC) in the following, while the remaining qubits are kept in a reference (separable) configuration.
Quantum computation proceeds through sequences of control pulses that operate collectively on one or more of three species of qubits. These pulses serve to rigidly shift the position of the ICC  in the ladder and activate single- and two-qubit gates on its elements at specific locations in the device, both enabled by the {\it blockade interactions} that connect the qubits. In our device, such interactions are implemented via the unwanted ZZ coupling mentioned above. As we will discuss below, by exploiting the fact that ZZ interactions can alleviate some of the degeneracies in the energy spectra of neighboring superconducting qubits, we selectively impede specific energy transitions, effectively emulating Rydberg blockade~\cite{cesa2023universal} on a superconducting platform.

\paragraph{Model.} We propose the quantum computing architecture depicted in Fig.~\ref{fig:transarray}, which consists of a 2D ladder with $N$ parallel rows. The ladder contains 
$2N+3$  columns, each housing a distinct type of superconducting qubits chosen among three
possible species $A$, $B$, and $C$ (represented, respectively, by red, blue, and green squares in the figure), organized
in the
\begin{equation}\label{CABA} 
\underbrace{CABA}_{}\underbrace{CABA}_{}\underbrace{CABA}_{}\cdots   
\end{equation}
pattern. The first (last) two columns on the left (right) boundary of the ladder serve as dedicated areas for the initialization (read-out) stage of quantum computation. The central portion of the scheme, encompassing $2N-1$ columns, represents instead the processing area of the model where the quantum computation takes place.
In this area, starting from the third column, each of the $B$- and $C$-type columns also incorporates an additional $A$-type qubit. This $A$-type qubit acts as a ``coupler" between adjacent rows of the ladder, enabling two-qubit logic gates~\cite{NUMBER}.
All qubits within a group share the same level spacing: $\hbar\omega_A$ for $A$-type qubits, $\hbar\omega_B$ for $B$-type qubits, and $\hbar\omega_C$ for $C$-type qubits. Exceptions to this rule, marked with a black circle or triangle in Fig.~\ref{fig:transarray}, have level spacings slightly detuned from the nominal values to compensate for the ``anomalous'' (i.e.,~different) number of nearest-neighboring 
ZZ interacting elements  (see the next paragraphs for details). 
In our model, qubits within the same row are interconnected via nearest-neighbor ZZ coupling, with uniform strength $\hbar \zeta$, depicted as black springs in the figure. This interaction also links additional intrarow coupler $A$-type qubits with adjacent $B$ or $C$ elements in the same column.
Crucially, all $A$-type qubits are collectively driven by a {\it single} time-dependent external electrical signal, $V_A(t)$, via a continuous red line. Similarly, all $B$-type [$C$-type] qubits, including those with a black circle or triangle, are collectively driven by signal $V_B(t)$ [$V_C(t)$] through a dedicated blue [green] line. 
Finally, the crossed qubits in the device identify a subset of superconducting qubits $A$, $B$, or $C$ that, while maintaining their nominal level spacings, couple with the external electrical signals $V_A(t)$, $V_B(t)$, or $V_C(t)$  at augmented Rabi frequencies. As shown in the figure, crossed elements of the same type never appear in the same row or column. Additionally, the pattern~(\ref{CABA}) ensures that any two columns that contain crossed elements of the same type $B$ or $C$ are always separated by a sequence of  {\it three}  columns  that contain no crossed qubits of that same type. We point out that this feature, which is crucial for implementing single-qubit and two-qubit gates,
can be engineered {\it at the level of circuit fabrication} and needs no extra control (see the Supplemental Material, SM~\cite{supplemental}). 

The total Hamiltonian of the ladder in Fig.~\ref{fig:transarray} can be written as $\hat{H}(t): = \hat{H}_0 + \hat{H}_{\rm drive}(t)$, where
\begin{equation}\label{Htot}
\hat{H}_0 := \sum_{\chi \in \{A,B,C\}} \sum_{i \in \chi} \frac{\hbar \omega_i}{2} \hat{\sigma}^{(z)}_{i}
+  \sum_{\langle i,j \rangle} \frac{\hbar \zeta}{2} \hat{\sigma}^{(z)}_{i} \otimes \hat{\sigma}^{(z)}_{j} 
\end{equation}
describes the local energy contribution of the superconducting qubits and their always-on ZZ interactions, while
\begin{equation}\label{Hdrive} 
\hat{H}_{\rm drive}(t) := \sum_{\chi \in \{A,B,C\}} \sum_{i \in \chi} \hbar \Omega_\chi(t)\sin(\omega_{\mathrm{d},\chi}t + \phi_{\chi}(t)) \hat{\sigma}^{(y)}_i
\end{equation} 
represents the time-dependent driving owing to the classical control lines.

In Eqs.~(\ref{Htot}) and~(\ref{Hdrive}), $\hat{\sigma}^{(x,y,z)}_{i}$ represent the Pauli matrices acting on the Hilbert space of the $i$th qubit, expressed in the local energy basis ${\vert g_i \rangle := (0,1)^{\rm T}, \vert e_i \rangle} := (1,0)^{\rm T}$. The summation in the second term on the right-hand side of Eq.~(\ref{Htot}) runs over all nearest-neighbor qubits. The parameter $\omega_{\mathrm{d},\chi}$ denotes the oscillation frequency of the driving pulse $V_\chi(t)$, while $\Omega_\chi(t)$ and $\phi_\chi(t)$ define the time-dependent Rabi frequency and phase of such control. 
For simplicity, in writing Eq.~(\ref{Htot}), we have not specified that when $i$ identifies a blue [green] qubit with a black circle, the corresponding qubit's level spacing becomes $\hbar(\omega_B-\zeta)$  [resp. $\hbar(\omega_C-\zeta)$] instead of $\hbar\omega_B$ [$\hbar\omega_C$], and when it identifies a blue [green] qubit with a black triangle, it becomes $\hbar(\omega_B+\zeta)$ 
[resp. $\hbar(\omega_C+\zeta)$]
instead of $\hbar\omega_B$ [$\hbar\omega_C$]. Similarly, whenever in Eq.~(\ref{Hdrive}) $i$ identifies a crossed qubit, $\Omega_{\chi}(t)$ is replaced by $2\Omega_\chi(t)$.
Apart from these local adjustments, which must be implemented once and for all at fabrication level, it is important to note that $\Omega_\chi(t)$, $\phi_\chi(t)$, and $\omega_{\mathrm{d},\chi}$ are independent of the index $i$, indicating that they represent a control acting {\it globally} on all qubits of type $\chi$.

\begin{figure}[t!]
\centering
\includegraphics[width=1.0\columnwidth]{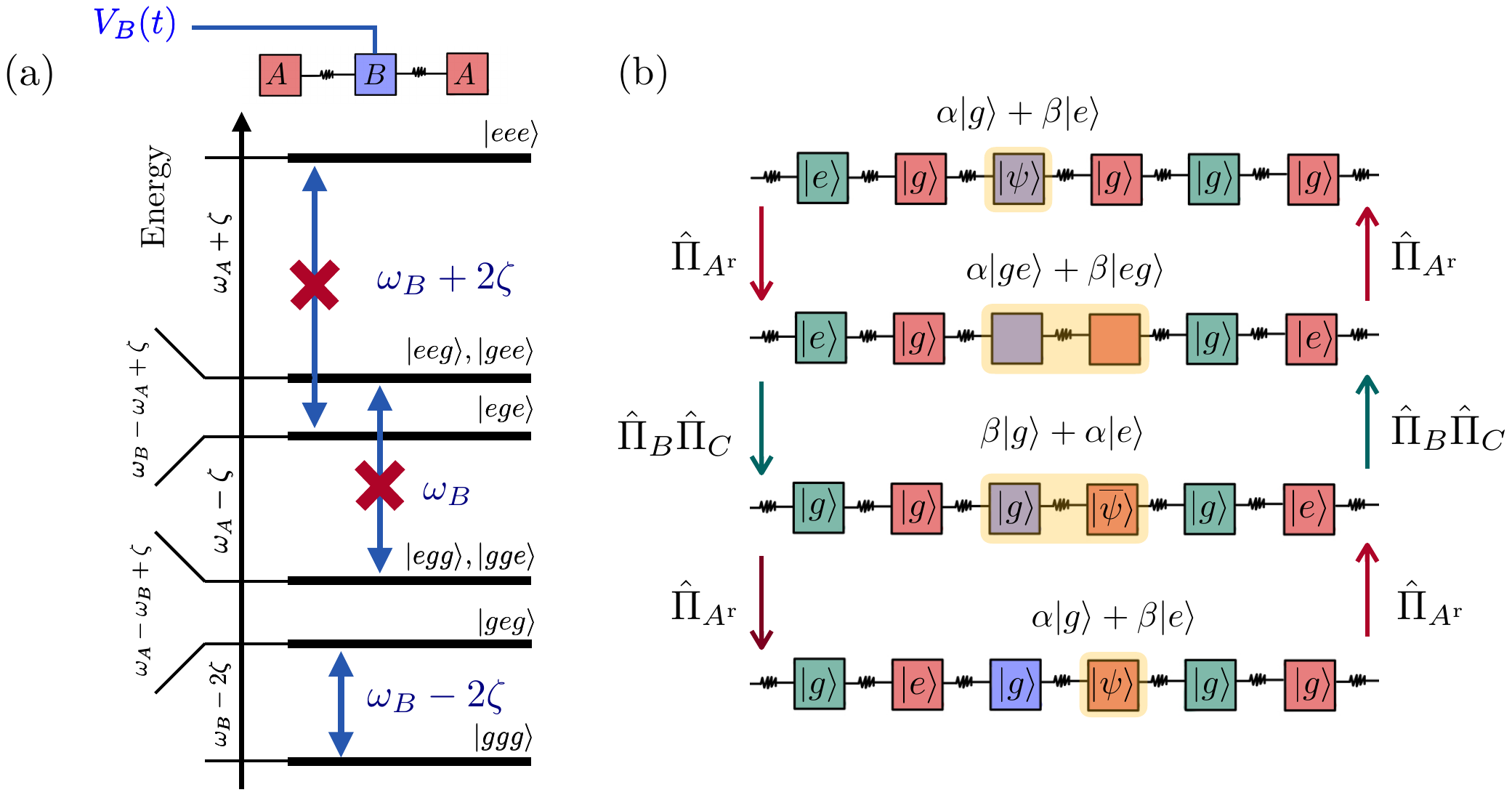}
\caption{(a) Energy spectrum of three qubits 
$ABA$ coupled via nearest-neighbor ZZ couplings as in $\hat{H}_0$: a driving pulse
of frequency $\omega_{\mathrm{d},B} = \omega_{B} - 2\zeta$ acting on the central $B$-type element can only activate transitions between the two lowest energy levels. The energy spectrum for the qubit triplet $ACA$ is analogous to the one shown. In this case, the driving pulse must have a frequency $\omega_{\mathrm{d},C} = \omega_{C} - 2\zeta$. The same argument holds for the $CAB$ triplet ($C$- and $B$-type qubits are driven at the same time thus they can be seen as the same species in this context) where the driving pulse is now $\omega_{\mathrm{d},A} = \omega_{A} - 2\zeta$. A generalization of this scheme for a $ABACA$ qubit chain is shown in the SM~\cite{supplemental}.
(b) Analysis of the sequence $\hat{U}_{\rm shift}= \hat{\Pi}_{A^{\rm r}}
\hat{\Pi}_{B}\hat{\Pi}_{C}
\hat{\Pi}_{A^{\rm r}}$ 
that induces a collective motion of the interface 
for the case of a single ($N=1$) row. 
\label{fig:transitions}}
\end{figure}

\paragraph{Blockade regime via ZZ interactions.}  
Throughout our analysis, we assume
that two ZZ-coupled nearest-neighbor qubits
are never simultaneously driven~\cite{DEFIMPULS}. 
Furthermore, in the time intervals where the control $V_{\chi}(t)$  is active, we shall assume the corresponding $\Omega_\chi(t)$ and $\phi_\chi(t)$ values to be constant.
Under these driving conditions, there exists a particular choice of $\omega_{\mathrm{d},\chi}$, which allows us to implement a dynamical blockade regime. 
A physical insight into the origin of this effect
can be obtained considering the energy spectrum of the
Hamiltonian $\hat{H}_0$ for the simplified case where one has three alternating qubits, say the $ABA$ triplet shown in
Fig.~\ref{fig:transitions}(a), arranged in a single row. A generalization of Fig.~\ref{fig:transitions}(a) to a qubit chain $ABACA$ is provided in the SM~\cite{supplemental}. In the $ABA$ case, which capture the essential physics, one notices that, in the rotating wave approximation (RWA), if we apply   
a pulse $V_{B}(t)$ of frequency $\omega_{\mathrm{d},B} := \omega_{B} - 2\zeta$ to the central element, only the transition between states $|ggg\rangle$ and $|geg\rangle$ is activated, while other transitions are not allowed since they are off-resonance.
In other words, whenever at least one of the ``external'' qubits of the triplet resides in the excited state $|e\rangle$, the system remains unaffected by the specified control pulse~\cite{g_AB}. 
A generalization of this result to the full ladder
can be obtained by describing the system's evolution in the time-dependent reference frame defined by the controls. In particular, it is possible to show (see the SM~\cite{supplemental}) that applying the unitary transformation $\hat{U}_{\rm rf}(t) := \bigotimes_i e^{i\hat{\sigma}_i^{(z)} \omega_{\mathrm{d,}i}t/2}$ in the realm of the RWA, the Hamiltonian of the model reduces to
\begin{eqnarray}\label{Hfinal}
\hat{H}_{\text{rf}}(t)&\simeq& \sum_{\chi \in \{A,B,C\}} \sum_{i \in \chi} \frac{\hbar \Omega_\chi(t)}{2} \Big[e^{i\phi_{\chi}(t)}\vert g_i \rangle \langle e_i \vert + {\rm H.c.} \Big]  \nonumber \\
&&+  \sum_{\langle i,j \rangle} 2 \hbar \zeta \vert e_i e_{j} \rangle \langle e_i e_{j} \vert~.
\end{eqnarray}
If the dimensionless parameter $\eta_{\rm BR}:=|\zeta/ \Omega_{\chi}|\gg 1$ (i.e.~if the ZZ coupling is the largest energy scale), $\hat{H}_{\text{rf}}(t)$ exactly mimics  the effective Rydberg-blockade Hamiltonian employed in Ref.~\cite{cesa2023universal}. 
We here stress that this Hamiltonian applies to {\it all} the qubits of the scheme, including  the off-detuned elements  indicated by  black triangles or circles in 
Fig.~\ref{fig:transarray}. 

\paragraph{Dynamics and quantum computation.}

The evolution induced by $\hat{H}_{\text{rf}}(t)$   can be expressed as ordered products of the form 
\begin{eqnarray} \label{string} 
\hat{U}_{\rm tot} = \hat{U}_{\chi_{\ell}}^{(\ell)} \cdots \hat{U}_{\chi_2}^{(2)} \hat{U}_{\chi_1}^{(1)}~, 
\end{eqnarray}
where $\hat{U}_{\chi_{\ell}}^{(\ell)}$ is the unitary operator associated with the $\ell$th time window, where $V_{\chi_{\ell}}(t)$ is active. Each unitary operator $\hat{U}_{\chi}$ can be described as $\hat{U}_{\chi} := e^{- i \hat{H}_{\chi} \tau_{\chi} / \hbar}$, with $\hat{H}_{\chi}=\hat{H}_{\text{rf}}(\tau_{\chi})$. Restricting ourselves to the blockade $\eta_{\rm BR}\gg 1$ regime,  we can consider the interaction term of $\hat{H}_{\chi}$  as a constraint 
that prevents any dynamical evolution on $\chi$-type qubits unless
both of their nearest-neighbor $\overline{\chi}$-type qubits~\cite{bar_chi} are in the ground state $|g\rangle$. This implies that Eq.~\eqref{string}
reduces  to a  product of 
identical {\it control-unitary} operations~\cite{Nielsen2010}, where each qubit of the $\chi$ type 
is controlled by its first neighboring sites 
of $\overline{\chi}$ type.
One can prove (see the SM~\cite{supplemental}) that, by using sequences~\eqref{string} that only involve a finite set of $\hat{U}_\chi$ activated by 
the same control line $V_{\chi}(t)$, we can induce arbitrary evolutions of the form
\begin{equation}\label{ccugen} 
\hat{W}_{\chi}(\theta',\bm{n}'; \theta'',\bm{n}'')
:=  \hat{W}_{\chi^{\rm r}}(\theta', \bm{n}') \hat{W}_{\chi^{\times}}(\theta'', \bm{n}'')~,
\end{equation}
where $\chi^{\rm r}$ ($\chi^{\times}$) is the subset of  $\chi$ that includes all its regular (crossed) elements. For
$\xi\in \{  \chi^{\rm r},\chi^{\times}\}$,  
$\hat{W}_{\xi}(\theta, \bm{n})$
is a control-unitary transformation that applies to all qubits in $\xi$, a uniform, single-qubit rotation
$\hat{\mathbb{R}}_i(\theta,\bm{n})
:= e^{ -i \frac{\theta}{2} \bm{n} \cdot \vec{\sigma}_{i}}$ 
parametrized by the 3D unit vector $\bm{n}$ and the angle $\theta\in [0,2\pi]$.
Of particular relevance  are the unitaries~(\ref{ccugen})
where only one 
of the parameters $\theta'$, $\theta''$ differs from zero, which correspond to transformations
that  selectively operate on either~$\chi^{\rm r}$ or $\chi^{\times}$. This allows for the individual addressing of both regular and crossed qubits using only global pulses.

Following Ref.~\cite{cesa2023universal}, 
we encode the quantum information on the 2D ladder by placing one logical qubit per row.
As indicated in Fig.~\ref{fig:transarray},
at each computational step, these qubits reside on a single  column  of the processing area of the device. Such ICC can be formed either by $A$-, $B$-, or $C$-type qubits. However, if the ICC is formed by $B$-
or $C$-type qubits,  the red-crossed elements in Fig.~\ref{fig:transarray}, which act as intrarow couplers, 
are not involved in the encoding but will always remain into the ground state $|g\rangle$ of their local Hamiltonian. 
 
The ICC is positioned at the interface between two strings of qubits: a “ferromagnetic" phase ($\vert g g g g\dots\rangle$) on the right, and a “paramagnetic" phase ($\vert \cdots g e g e g \rangle$) on the left. Assuming that at a given computational step a (possibly entangled) state $|\Psi\rangle$ 
of the $N$-logical qubit  is
located in the $k$th  column of the ladder, the global quantum state of the model is described by {\it well-formed} vectors of the form
\begin{eqnarray}\label{coding}
&& |\Psi; {k}\rangle : =\Big( \cdots  |g^{\otimes N} \rangle_{k-3} \otimes |e^{\otimes N}\rangle_{k-2} \otimes |g^{\otimes N}\rangle_{k-1} \Big) 
\\ && \quad\;\; \otimes
|\Psi\rangle_k   \otimes 
\Big(   |g^{\otimes N} \rangle_{k+1}  \otimes  |g^{\otimes N} \rangle_{k+2} \otimes 
|g^{\otimes N} \rangle_{k+3}    \cdots \Big)~,
\nonumber
\end{eqnarray}
where, for $k=1,\cdots, 2N+3$, the ket $|\cdots\rangle_{k}$ refers to the state of the $k$-th column of the device. As shown in the SM~\cite{supplemental}, starting from a fully ferromagnetic phase, where all the qubits of the device are in the ground state $|g\rangle$, a specific control pulse $V_{C}(t)$ can promote the first $C$-type column of the initialization area of Fig.~\ref{fig:transarray} in $|e^{\otimes N}\rangle$. Formally, this  produces a vector $|\Psi; {k}\rangle$ where the ICC is located in the first $B$-type column
of the processing area, and a logical state $|\Psi\rangle:= |g^{\otimes N}\rangle$.  Thanks to Eq.~(\ref{coding}), one can show that, under 
$e$--$e$ blockade conditions, there exist
control unitary evolutions~(\ref{string})  that 
enable universal QC. 
The key ingredients that make this feasible are: 
(a)~the ability to {\it rigidly} move the ICC at any position in the ladder (including the last column of the read-out area where measurements can be performed at the end of the computation); (b)~the possibility to implement arbitrary single-qubit gates on the crossed element of 
the $B$- or $C$-type column where the ICC 
is located, while leaving the rest of the qubits unaffected; and (c)~the ability to activate a nontrivial two-qubit entangling gate (e.g.,~a control-Z operation) between the $j$th and $(j+1)$th logical qubits when the ICC is located in the $B$- or $C$-type column that contains the crossed $A$-type qubit, which acts as a coupler between the $j$th and $(j+1)$th rows of the ladder.

To achieve task (a), we  identify a sequence $\hat{U}_{\rm shift}$~\cite{U_shift} of transformations
(\ref{ccugen}) 
that  shifts the horizontal position of the ICC
without affecting its internal state,  i.e.,
$\hat{U}_{\rm shift} |\Psi; {k}\rangle =  |\Psi; {k + 1}\rangle$, see Fig.~\ref{fig:transitions}(b). 

In the case in which the ICC is located in a $\chi$-type column
with  $\chi\in\{B,C\}$, the 
operations needed for implementing the single-qubit gates of point (b) can be realized by composing
sequences of the form
$\hat{Z}_{A^{\rm r}}^{{\rm tot}} \hat{W}_{\chi^{\times}}(\theta/2,-\bm{n}_\perp) 
\hat{Z}_{A^{\rm r}}^{{\rm tot}} \hat{W}_{\chi^{\times}}(\theta/2,\bm{n}_\perp)$ with $\bm{n}_\perp$ orthogonal to $\bm{z}$ and 
$\hat{Z}_{A^{\rm r}}^{{\rm tot}}$ obtained from (\ref{ccugen}) for $\theta'=2\pi$ and $\theta''=0$~ (see the\cite{supplemental}). 
Indeed, thanks to the fact that the crossed $\chi$-type qubits are located on columns, which are at least three columns apart from each other, when acting on $|\Psi; {k}\rangle$ the above transformation will effectively correspond to the application of the single-qubit rotation  $\hat{\mathbb{R}}(\theta,\bm{n}_\perp)$ on the crossed element of the ICC~\cite{single-q}. Finally, once the ICC is aligned with  a two-qubit gate, we only need to use a single $\hat{Z}_{A^{\times}}^{\rm tot}$ evolution obtained by setting $\theta'=0$ and $\theta''=2\pi$ in Eq.~(\ref{ccugen}), which leaves unchanged every regular $A$-type qubit and induces conditional-phase shift~\cite{Nielsen2010} on the crossed ones.

\paragraph{Discussion.}  
Exploiting ZZ interactions,
we achieved universal QC on $N$ logical qubits through external {\it global} pulses that act on a ladder of
$N_{\rm tot} = 2N^2+4N-1$ physical qubits~\cite{NUMBER}.
In contrast with the Rydberg-atom proposal~\cite{cesa2023universal}, which relies on the use of two different species of
physical qubits,  our scheme uses three species. This choice
allows us to roughly halve the total number of physical qubits of the device, which for the two-species model~\cite{cesa2023universal}  scales as  $N_{\rm{tot}}\simeq 4N^2$ for $N\gg 1$. As discussed above, the condition $\eta_{\rm BR}\gg 1$ is crucial for our device's operation. Numerical checks in the SM~\cite{supplemental} confirm that $\zeta$ must be at least on the order of $5 \Omega_\chi$. The largest values of $\zeta$ that have been experimentally achieved~\cite{DiCarlo2009} are on the order of $100~{\rm MHz}$, while typical Rabi frequencies can reach values as small as $5~{\rm MHz}$~\cite{Sung2021}, suggesting that the blockade regime is fully within reach in current superconducting-qubit platforms. 

In order to achieve a fault-tolerant quantum computation, error-correction (EC) schemes are necessary. Since our machine is universal, any standard fault-correction procedure can be implemented. In particular, regarding, e.g.,~surface codes~\cite{surface-code} or qLDPC codes~\cite{LDPC1, LDPC2}, we mention that while our proposal uses only nearest-neighbor couplings, nothing prevents us from introducing long-range couplings between logical qubits in order to reach 2D connectivity requirements. This can be experimentally carried out, for example, by following the approaches of Refs.~\cite{Marxer2023, Deng2024}. 
We finally mention that existing EC schemes for globally controlled quantum computers have been studied in the past~\cite{Bririd2004, Kay2005, Kay2007, Fitzsimons2007, Fitzsimons2009}; however, globally driven EC codes for our machine will be subject of future research.
Moreover, a significant challenge of our architecture is the impact of inhomogeneities in SC qubit frequencies and coupling strengths on protocol fidelity. Although a detailed noise analysis is beyond the scope of this Letter, we remark that systematic fabrication errors could be treated as standard dynamical errors within error correcting schemes, as shown in Ref.~\cite{Aharonov1999}, thus allowing the strategies discussed above to be applied. An alternative strategy involves a {\it hybrid} “bottom-up mitigation approach", where global control is applied to small qubit patches or chiplets (with fabrication inhomogeneities below a set tolerance). These patches will be then connected to each other~\cite{Gold2021, Niu2023}. For instance, suppose that in order to have $N$ logical qubits, we need to build $k$ patches with inhomogeneities below the above mentioned tolerance, each encoding $M$ logical qubits. For each patch we have $\mathcal{O}(1)$ control lines. By coupling the $k$ chiplets we still have $\mathcal{O}(k)$ wires with $k\sim N/M$. This configuration would still significantly reduce the number of control lines required for universal QC. The calculation of the tolerance required is left for future work.
Another potential challenge is that qubit state readout is currently limited to the rightmost column of the ladder, complicating the individual qubit tune-up typically required for gate calibration. To address this, flip-chip technology could be employed, for example by coupling the computational chip to a top readout chip that performs individual readout of the qubits inside the ladder~\cite{Rosenberg2017}.

\paragraph{Acknowledgments.}
We thank F. Cesa and M. Riccardi for useful discussions, and J. Despres for useful comments on the draft. This work has been funded by the European Union - NextGenerationEU, Mission 4, Component 2, under the Italian Ministry of University and Research (MUR) Ex- tended Partnership PE00000023 National Quantum Science and Technology Institute -- NQSTI -- CUP J13C22000680006. 

M.P. and V.G. are co-founders and shareholders of Planckian.
At the time of their contributions, authors affiliated with Planckian are either employees at Planckian or PhD students collaborating with Planckian.


\begin{thebibliography}{99}
\bibitem{NUMBER} 
The total number of row-couplers  is equal to $N-1$, bringing the 
total number of superconducting qubits  to $N_{\rm{tot}}=(2N+3)N+ N-1=2N^2+4N-1$, while the total number of columns is $N_{\rm column}=2N+3$.
\bibitem{DEFIMPULS} More specifically  at any given time, either the control pulse  $V_A(t)$ is active while  $V_B(t)$ and  $V_C(t)$ are not, or, if either  $V_B(t)$ or  $V_C(t)$ is active, then  $V_A(t)$ is not
i.e. $\Omega_A(t)\neq 0 \Rightarrow \Omega_B(t)=\Omega_C(t)=0$, and 
$\Omega_B(t)\neq 0$ or $\Omega_C(t)\neq 0$ $\Rightarrow \Omega_A(t)=0$.
\bibitem{bar_chi}
Here, the symbol $\overline{\chi}$ denotes a species that is distinct from $\chi$.
\bibitem{single-q} 
The key observation here is that when the $k$th column of  $|\Psi; {k}\rangle$ is
made of $B$-type qubits, the unitary 
$\hat{Z}_{A^{\rm r}}^{{\rm tot}}$ defined at the end of the previous section is equivalent to the application of a $\hat{\sigma}^{(z)}$ gate  to each of the element of the ICC, i.e., $\hat{Z}_{A^{\rm r}}^{{\rm tot}}|\Psi; {k}\rangle=
| \Big(\hat{\sigma}^{(z)}\Big)^{\otimes N}\Psi; {k}\rangle$. 
Notably, the same property holds also when we replace the $|g^{\otimes N}\rangle$ 
state of  a
ferromagnetic $B$-type  column of  $|\Psi; {k}\rangle$ with a generic $N$-qubit state. Exploiting this effect it then follows that, 
given a unit vector $\bm{n}_\perp$ orthogonal to $\bm{z}$, the sequence 
$\hat{Z}_{A^{\rm r}}^{{\rm tot}} \hat{W}_{B^{\times}}(\theta/2,-\bm{n}_\perp) 
\hat{Z}_{A^{\rm r}}^{{\rm tot}} \hat{W}_{B^{\times}}(\theta/2,\bm{n}_\perp)$ maps $|\Psi; {k}\rangle$ into 
a new state $|\Psi'; {k}\rangle$ with a new ICC vector  $|\Psi'\rangle$ obtained from $|\Psi\rangle$ via the application of the
single-qubit rotation $\hat{\sigma}^{(z)}\hat{\mathbb{R}}(\theta/2,-\bm{n}_\perp)\hat{\sigma}^{(z)}\hat{\mathbb{R}}(\theta/2,\bm{n}_\perp)=\hat{\mathbb{R}}(\theta,\bm{n}_\perp)$ on the crossed element of the column.
Analogous considerations apply also to the case in which the ICC is of $C$ type.
\end{thebibliography}
\end{document}


\title{\Large Supplemental Material for: \\ ``Globally driven superconducting quantum computing architecture''}
\author{Roberto Menta}
\affiliation{Planckian srl, I-56127 Pisa, Italy}
\affiliation{NEST, Scuola Normale Superiore, I-56127 Pisa, Italy}
\author{Francesco Cioni}
\affiliation{NEST, Scuola Normale Superiore, I-56127 Pisa, Italy}
\author{Riccardo Aiudi}
\affiliation{Planckian srl, I-56127 Pisa, Italy}
\author{Marco Polini}
\affiliation{Planckian srl, I-56127 Pisa, Italy}
\affiliation{Dipartimento di Fisica dell’Universit\`{a} di Pisa, Largo Bruno Pontecorvo 3, I-56127 Pisa, Italy}
\author{Vittorio Giovannetti}
\affiliation{Planckian srl, I-56127 Pisa, Italy}
\affiliation{NEST, Scuola Normale Superiore, I-56127 Pisa, Italy}

\maketitle

In this Supplemental Material file we provide additional technical details in order to clarify the results presented in the main text. In Section~\ref{sec:cQED toolbox}, we briefly summarize well-known facts about superconducting qubits and their control via external classical drivings. We also discuss how to couple two superconducting qubits via the longitudinal ZZ coupling and how such interaction, in addition to off-resonance pulses, can induce the $e$--$e$ blockade regime discussed in the main text. We also briefly recap available experimental paths to realize ZZ interactions, as available in the literature. In Section~\ref{sezDYN}  we solve the dynamical evolution of the ladder 
in the rotating frame, deep in the $e$--$e$ blockade regime. In particular, we provide an explicit derivation of Eq.~(6) in the main text. In Section~\ref{secENC} we present a detailed analysis of how to perform universal quantum computation in our architecture. Finally, in Section~\ref{sec:Numerical Simulations} we present basic numerical simulations of our pulsed protocol for globally-driven universal quantum computation.

\section{\label{sec:cQED toolbox}Fundamentals of superconducting qubits}

In this Section, with the sole aim of producing a self-contained document, we briefly recap textbook fundamentals of superconducting qubits.

\begin{figure}[t]
\centering
\includegraphics[scale=0.35]{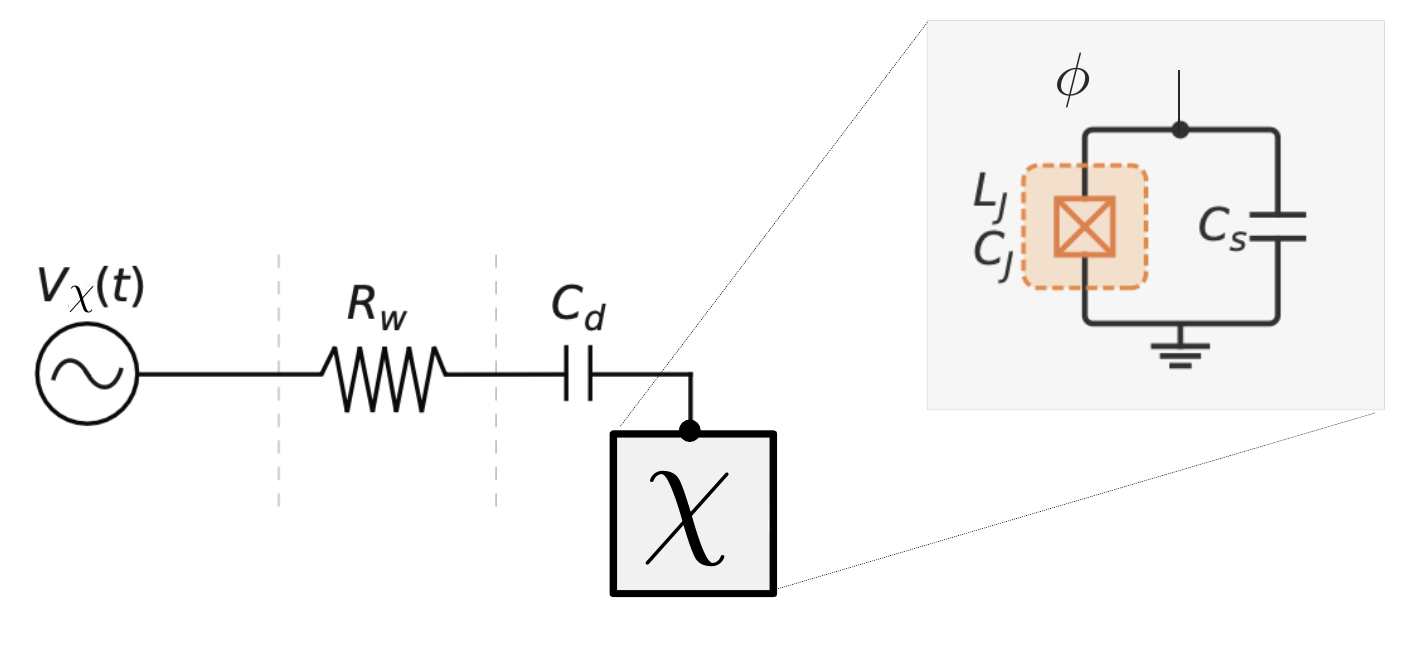}
\caption{\label{fig:driven-qubit} Superconducting qubit driven by an external electric potential $V_{\chi}(t)$. For a crossed superconducting qubit (refer to Fig.~1 in the main text), the driving capacitance is doubled compared to normal $\chi$-type qubits, i.e. $\mathrm{C_d} \rightarrow 2\mathrm{C_d}$, resulting in $\Omega_{\chi} \rightarrow 2\Omega_{\chi}$. That is essential for implementing quantum gates in our architecture.}
\end{figure}

A superconducting qubit~\cite{krantz2019quantum, blais2021circuit} is a superconducting circuit that can be effectively described as a two-level system. The Hamiltonian is
\begin{equation}\label{H-qubit}
    \hat{H}_0 = \dfrac{\hbar \omega_\chi}{2} \hat{\sigma}^{(z)}~,
\end{equation}
where $\chi = A,B,C $ denotes the qubit type, $\hat{\sigma}^{(z)}$ is the Pauli-z operator and $\omega_{\chi} = (E_e - E_g)/\hbar$ is the energy splitting between the excited $\vert e \rangle := (1,0)^{\rm T}$ and ground state $\vert g \rangle := (0,1)^{\rm T}$ of the qubit. The magnitude of  $\omega_{\chi}$ depends only on the geometric parameters of the quantum chip. Indeed, for a simple qubit design as the one shown in Fig.~\ref{fig:driven-qubit}, the frequency is given by $\omega_\chi = (\sqrt{8E_\mathrm{J}E_\mathrm{C}} - E_\mathrm{C})/\hbar$ where $E_\mathrm{C} = e^2/(2\mathrm{C}_{\Sigma})$ is the total charging energy, $\mathrm{C}_{\Sigma} = \mathrm{C}_\mathrm{s} + \mathrm{C}_\mathrm{J}$ is the total capacitance, including the shunt capacitance $\mathrm{C}_\mathrm{s}$ and the self-capacitance of the junction $\mathrm{C}_\mathrm{J}$, while $E_\mathrm{J} = \mathrm{I}_\mathrm{c} \Phi_0/2\pi$ is the Josephson energy, with $\mathrm{I_c}$ being the critical current of the junction and $\Phi_0=h/(2e)$ being the superconducting magnetic flux quantum. It turns out that there are some kind of superconducting qubits known as {\it split transmons} (SQUID), which enable frequency tunability using externally applied magnetic field \cite{krantz2019quantum}.
By connecting the $\chi$-type superconducting qubit to an external electric potential source  $V_{\chi}(t)$, as shown in Fig.~\ref{fig:driven-qubit}, we are able to perform unitary rotations on the qubit. Here we outline how this is achieved. 
Following the connection with the external classical source, we introduce an additional interaction term in the Hamiltonian, i.e.
\begin{equation}\label{H-drive}
    \hat{H}_{\text{drive},\chi}(t) = \mathcal{V} V_{\chi}(t) \;  \hat{\sigma}^{(y)} \ ,
\end{equation}
where $\mathcal{V} := (\mathrm{C_d}/\mathrm{C}_{\Sigma}) \sqrt{\hbar/(2\sqrt{\mathrm{L_J/C_s}})}$ is a constant depending on the construction parameters of the circuit. Here, $\mathrm{L_J}$ is the inductance associated to the Josephson Junction. We can generally assume that the time-dependent part of the voltage takes a generic oscillatory form,
\begin{equation}
        V_{\chi}(t) = \mathcal{S}_{\chi}(t)  \sin(\omega_{\rm{d},\chi} t + \phi_{\chi}(t)) \;, 
        \end{equation} 
where $\omega_{\rm{d},\chi}$ is the fundamental frequency of the driving pulse, $\mathcal{S}_{\chi}(t)$ is an envelope function, and finally $\phi_{\chi}(t)$ is the phase of the driving that in the following we shall assume to be time-dependent. 
It is instructive to move into a frame rotating with the drive frequency $\omega_{\rm{d},\chi}$. This is done by defining the unitary operator $\hat{U}_{\text{rf}}(t) = \exp ({i\omega_{\rm{d},\chi}\hat{\sigma}^{(z)} t/2})$ such that the new state in the rotating frame is $\vert \psi_{\text{rf}} (t) \rangle = \hat{U}_{\text{rf}}(t) \vert \psi(t) \rangle$. The time-evolution in this new frame is given by the Schr\"odinger equation,
\begin{equation}
    i \hbar \dfrac{\partial}{\partial t} \vert \psi_{\text{rf}} (t) \rangle = \Big( \underbrace{i \hbar \dot{\hat{U}}_{\text{rf}}(t) \hat{U}_{\text{rf}}^{\dagger}(t)+ \hat{U}_{\text{rf}}(t) \hat{H}_0 \hat{U}_{\text{rf}}^{\dagger}(t)}_{\hat{H}'_{0}}
    +  \underbrace{\hat{U}_{\text{rf}}(t)\hat{H}_{\text{drive},\chi}(t) \hat{U}_{\text{rf}}^{\dagger}(t)}_{ \hat{H}'_{\text{drive},\chi}(t)}
 \Big) \vert \psi_{\text{rf}} (t) \rangle \ .
\end{equation}
Simple manipulations show that in this frame we have ${\hat{H}}'_0 = \hbar\Delta_\omega\hat{\sigma}^{(z)}/2$, where $\Delta_\omega = \omega_\chi - \omega_{\rm{d},\chi}$ is the detuning. 
The driving term instead yields 
\begin{eqnarray}
  \hat{H}'_{\text{drive},\chi}(t) &=& \mathcal{V} V_{\chi}(t) (\cos(\omega_{\rm{d},\chi} t)  \hat{\sigma}^{(y)} + \sin(\omega_{\rm{d},\chi} t) \hat{\sigma}^{(x)}) \nonumber 
  \\ 
  &=& \nonumber 
 {\hbar  \Omega_{\chi}(t)} \sin(\omega_{\rm{d},\chi} t + \phi_{\chi}(t))  (\cos(\omega_{\rm{d},\chi} t)  \hat{\sigma}^{(y)} + \sin(\omega_{\rm{d},\chi} t)\hat{\sigma}^{(x)})\\ 
  &=&  \nonumber 
 {\hbar  \Omega_{\chi}(t)}\left(\frac{e^{i(\omega_{\rm{d},\chi} t + \phi_{\chi}(t))}-e^{-i(\omega_{\rm{d},\chi} t + \phi_{\chi}(t))}}{2i}
  \right)  \left(\frac{
 e^{-i\omega_{\rm{d},\chi} t} ( \hat{\sigma}^{(y)}+i\hat{\sigma}^{(x)})+
 e^{i\omega_{\rm{d},\chi} t} ( \hat{\sigma}^{(y)}-i\hat{\sigma}^{(x)})}{2}\right)\\
 &=&\nonumber 
{\hbar  \Omega_{\chi}(t)}\left(\frac{e^{i(\omega_{\rm{d},\chi} t + \phi_{\chi}(t))}-e^{-i(\omega_{\rm{d},\chi} t + \phi_{\chi}(t))}}{2}
  \right)  \left(
 e^{-i\omega_{\rm{d},\chi} t} |g\rangle\langle e|-
 e^{i\omega_{\rm{d},\chi} t} |e\rangle\langle g|\right)\;, 
\end{eqnarray}
with $\Omega_{\chi}(t) := \mathcal{V}  \mathcal{S}_\chi(t)/\hbar$  the time-dependent Rabi frequency. Notice that $\Omega_{\chi}(t)$ depends on $\mathcal{V}$, which itself depends on the capacitance coupling the qubit and the driving source. This allows us to change $\mathrm{C}_{\rm d}$ and consequently $\mathcal{V}$ in order to modify the time-independent part of the Rabi frequency. This procedure is the one adopted to realize crossed qubits (see the main text). Performing the multiplication and dropping fast rotating terms that will approximately average to zero (i.e. terms with $2 \omega_{\rm{d},\chi}$), known as rotating wave approximation (RWA), the Hamiltonian $\hat{H}'_{\text{drive},\chi}$ can be written as
\begin{equation}\label{H-driven-qubit}
     \hat{H}^{\prime(\text{RWA})}_{\text{drive},\chi}(t) =
     \frac{\hbar \Omega_{\chi}(t)}{2}\Big(e^{i\phi_{\chi}(t)}\vert g \rangle \langle e \vert + e^{-i\phi_{\chi}(t)} \vert e \rangle \langle g \vert \Big)\;, 
\end{equation}
which allows Rabi transitions between the ground and the excited states and vice-versa.

\subsection{The ZZ coupling implies the blockade regime }
In this Section, we show that in the rotating frame induced by the 
unitary transformation $\hat{U}_{\rm rf}(t) := \bigotimes_i e^{i\hat{\sigma}^{(z)}_i \omega_{\mathrm{d,}i}t/2}$, the Hamiltonian $\hat{H}(t)$ of the 2D ladder, under RWA reduces to 
${\hat{H}}_{\rm rf}(t)$ of~Eq.~(4) of the main text. 
Following the approach detailed in the previous Section moving in the rotating frame 
we have 
\begin{eqnarray} 
{\hat{H}}_{\rm rf}(t) = \label{newHM} 
 \underbrace{i \hbar
\dot{\hat{U}}_{\text{rf}}(t) \hat{U}_{\text{rf}}^{\dagger}(t) +  \hat{U}_{\text{rf}}(t) \hat{H}_0   \hat{U}_{\text{rf}}^{\dagger}(t)}_{\hat{H}'_{0}} + \underbrace{\hat{U}_{\text{rf}}(t) \hat{H}_{\rm drive}(t)    \hat{U}_{\text{rf}}^{\dagger}(t)}_{\hat{H}'_{\rm drive}(t)} \;, 
\end{eqnarray} 
with $\hat{H}_0$ and $\hat{H}_{\rm drive}(t)$ being the many-body Hamiltonians described in 
 Eqs.~(2) and (3) of the main text which we express here as
 \begin{eqnarray} \label{Htotsup}
\hat{H}_0 &=&  \sum_{i } \frac{\hbar \omega_i}{2} \hat{\sigma}^{(z)}_i
+  \sum_{\langle i,j \rangle} \frac{\hbar \zeta}{2} \hat{\sigma}^{(z)}_i \otimes \hat{\sigma}^{(z)}_j \;, \\
\hat{H}_{\rm drive}(t) &=&  \sum_{i } \hbar \Omega_\chi(t)\sin(\omega_{\mathrm{d},\chi}t +\phi_{\chi}(t)) \; \hat{\sigma}^{(y)}_i\;,\label{HdriveSUP} 
\end{eqnarray} 
without explicitly stating which type of qubit
is associated with the $i$-th qubit, but maintain the convention 
that if the index happens to identify 
a regular  $\chi$-type qubit then $\omega_i= \omega_{\chi}$, if instead identifies a $\chi$-type qubit with a black circle then $\omega_i=\omega_{\chi}^{\bullet}:=
 \omega_{\chi}-\zeta$, and finally if it identifies a 
$\chi$-type qubit with a black triangle then $\omega_i=
\omega_{\chi}^{\blacktriangle}:= \omega_{\chi}+\zeta$.

Notice hence that despite the presence of the ZZ couplings one has 
 $\hat{U}_{\text{rf}}(t) \hat{H}_0   \hat{U}_{\text{rf}}^{\dagger}(t)=\hat{H}_0$ and that 
 $ i \hbar
\dot{\hat{U}}_{\text{rf}}(t) \hat{U}_{\text{rf}}^{\dagger}(t)= -\sum_{i} \frac{\hbar \omega_{\mathrm{d,}i}}{2} \hat{\sigma}^{(z)}_i$. Accordingly 
the first contribution of Eq.~(\ref{newHM}) can be expressed as
\begin{equation} \label{H0prime}
\hat{H}'_0 := \frac{\hbar}{2}  \left( \sum_{i} ( \omega_i-\omega_{\mathrm{d,}i}) \hat{\sigma}^{(z)}_i
+  \sum_{\langle i,j \rangle}  \zeta \hat{\sigma}^{(z)}_i \otimes \hat{\sigma}^{(z)}_j\right) \;. \end{equation} 
Observe next that by imposing 
$\omega_{\mathrm{d,}\chi}=\omega_{\chi} - 2\zeta$, in the rotating frame the frequencies of the qubits becomes proportional to the number of ZZ connections they share with their neighbouring qubits, i.e. 
\begin{eqnarray} 
\left\{ \begin{array}{lcc}
\omega_\chi^{\bullet}-\omega_{\mathrm{d,}\chi} =  (\omega_{\chi}-\zeta)-(\omega_{\chi} - 2\zeta)=
\zeta && (\mbox{one connection}) \;,  \\
\omega_\chi-\omega_{\mathrm{d,}\chi} 
=  \omega_{\chi}-(\omega_{\chi} - 2\zeta) =
 2\zeta && (\mbox{two connections})\;,  \\
\omega_\chi^{\blacktriangle}-\omega_{\mathrm{d,}\chi} =
  (\omega_{\chi}+\zeta)-(\omega_{\chi} - 2\zeta)=
 3\zeta& & (\mbox{three connections}) \;.
\end{array} \right.
\end{eqnarray} 
Accordingly we can rewrite~(\ref{H0prime}) as a sum of triples of the form 
\begin{equation}\label{Htotnew}
\hat{H}'_0 =  \frac{\hbar \zeta}{2}  \sum_{\langle i,j\rangle}  
 \left(\hat{\sigma}^{(z)}_i+ \hat{\sigma}^{(z)}_i \otimes {\sigma}^{(z)}_j +
  {\sigma}^{(z)}_j\right) = \hbar 2\zeta  \sum_{\langle i,j\rangle}   |e_{i}e_{j}\rangle\langle e_{i}e_{j}| + {\rm const.} 
  \end{equation} 
  where in the last passage we used the identity 
\begin{eqnarray} 
\hat{\sigma}^{(z)}_i+ \hat{\sigma}^{(z)}_i \otimes {\sigma}^{(z)}_j +
  {\sigma}^{(z)}_j&=& 3 |e_ie_{j}\rangle\langle e_ie_{j}|- |e_ig_{j}\rangle\langle e_ig_{j}|-  |g_ie_{j}\rangle\langle g_ie_{j}|-  |g_ig_{j}\rangle\langle g_ig_{j}| \nonumber\\
  &=&4 |e_ie_{j}\rangle\langle e_ie_{j}| + {\rm const.} 
\end{eqnarray}
The driving term of Eq.~(\ref{newHM}) instead yields 
\begin{eqnarray}
\nonumber   \hat{H}'_{\text{drive}}(t)     &=&  \sum_i {\hbar  \Omega_{\chi}(t)} \sin(\omega_{\rm{d},\chi} t + \phi_{\chi}(t))  \Big(\cos(\omega_{\rm{d},\chi} t) \hat{\sigma}^{(y)}_i + \sin(\omega_{\rm{d},\chi} t)\hat{\sigma}^{(x)}_i\Big)\\
\nonumber 
  &=&  \sum_i {\hbar  \Omega_{\chi}(t)}\left(\frac{e^{i(\omega_{\rm{d},\chi} t + \phi_{\chi}(t))}-e^{-i(\omega_{\rm{d},\chi} t + \phi_{\chi}(t))}}{2i}
  \right)  \left(\frac{
 e^{-i\omega_{\rm{d},\chi} t} (\hat{\sigma}^{(y)}_i+i\hat{\sigma}^{(x)}_i)+
 e^{i\omega_{\rm{d},\chi} t} (\hat{\sigma}^{(y)}_i-i\hat{\sigma}^{(x)}_i)}{2}\right)\\
 &=&\sum_i {\hbar  \Omega_{\chi}(t)}\left(\frac{e^{i(\omega_{\rm{d},\chi} t + \phi_{\chi}(t))}-e^{-i(\omega_{\rm{d},\chi} t + \phi_{\chi}(t))}}{2}
  \right)  \left(
 e^{-i\omega_{\rm{d},\chi} t} |g_i\rangle\langle e_i|+
 e^{i\omega_{\rm{d},\chi} t} |e_i\rangle\langle g_i|\right)\;, 
 \end{eqnarray} 
 which under RWA becomes 
 \begin{eqnarray} 
   \hat{H}'_{\text{drive}}(t)  
 &\simeq&  \hat{H}^{\prime(\rm RWA)}_{\text{drive}}(t)  :=  \label{hdriveprime} 
 \sum_i \frac{\hbar \Omega_{\chi}(t)}{2}\left(
e^{i \phi_{\chi}(t)}|g_i\rangle\langle e_i| + e^{-i \phi_{\chi}(t)} |e_i\rangle\langle g_i| \right)\;.
\end{eqnarray}
Replacing~(\ref{Htotnew}) and~(\ref{hdriveprime}) into Eq.~(\ref{newHM})  gives~Eq.~(4) of the main text.

\subsection{Realizations of ZZ coupling}\label{sec:ZZ-realizations}
Our architecture relies on superconducting qubits that interact with their neighbors via a ZZ coupling. In modern quantum architectures, this kind of interaction is exploited to implement C-Phase gates and it has been already investigated in numerous experiments, using different strategies. In this Section, we present three of the possible ways this interaction can be realized.

\subsubsection{Cavity mediated coupling}
One way to reach an effective ZZ interaction between two superconducting qubits is to connect them dispersively through a cavity, i.e.~a waveguide whose resonance frequency is detuned with respect to the one of the qubits. This was experimentally demonstrated in Ref.~\cite{DiCarlo2009}, where they reached a ZZ interaction of magnitude 160 MHz. Such coupling can be obtained starting from the usual Tavis-Cummings Hamiltonian of two qubits interacting with the same cavity.
In the dispersive regime, where the cavity is far detuned with respect to the qubits, one can show that the cavity degrees of freedom can be eliminated via a Schrieffer-Wolff transformation. The result is an effective ZZ interaction of strength
\begin{equation}
    \zeta = - 2g^2_A g^2_B\left( \frac{1}{\delta_A\Delta_B^2}  + \frac{1}{\delta_B\Delta_A^2} + \frac{1}{\Delta_A\Delta_B^2} + \frac{1}{\Delta_A^2\Delta_B}   \right)\,,
    \label{ZZ_coupler}
\end{equation}
where $\Delta_\chi = \omega^{(\chi)}_{01} - \omega_{\rm c}$, $\delta_{\chi} = \omega^{(\chi)}_{01} - \omega^{(\chi)}_{12} $, with $\omega^{(\chi)}_{jk}$ being the level spacing between the states with $j$ and $k$ excitation for the $\chi$-type qubit with $\chi=A,B$; $g_A$ and $g_B$ correspond to the couplings between the corresponding qubit and the waveguide. The main issue of this scheme is that the cavity gives also rise to a SWAP interaction (of the type $\hat{\sigma}^{(A)}_+\hat{\sigma}^{(B)}_- + \hat{\sigma}^{(B)}_+\hat{\sigma}^{(A)}_-$) between the qubits. However, this can be neglected if one consider the qubits far detuned from each other.
    
\subsubsection{Qubit mediated coupling: Josephson junctions directly connected}
Another possibility for realizing the ZZ coupling is to place a third superconducting qubits (called coupler) that mediates the interaction between the two qubits. The coupler can be placed such that the Josephson junctions of the three qubits are directly connected to each other. The main advantage of this architecture is that it provides the ability to make the SWAP interaction vanishing for a certain coupler frequency. This scheme was implemented experimentally in Ref.~\cite{Kounalakis2018} and studied numerically in Ref.~\cite{Rasmussen2020}. The interaction engineered in this way depends on the circuit elements as
\begin{equation}
    \zeta = - \frac{E_{\rm J}^{\rm c} E_{\rm C}}{8\hbar E_{\rm J}}\,,
\end{equation}
with $E_{\rm J}$ and $E_{\rm C}$ being respectively the Josephson and charge energy of the qubits and $E_{\rm J}^{\rm c}$ the Josephson energy of the coupler.
Only small couplings have so far been measured ($<$ 10 MHz), but numerical calculations hint at the possibility of engineering stronger interactions up to 300 MHz, while at the same time keeping the SWAP interaction close to zero. However, this relies on a coupler qubit having a very low Josephson energy (20 MHz), which might not be feasible experimentally. Moreover, it could be challenging to scale this approach to a large number of qubits as the system becomes increasingly sensitive to flux noise, as it was pointed out in Ref.~\cite{Baker2022}.

\subsubsection{Qubit mediated coupling: Josephson junctions capacitively connected}
The last strategy we briefly mention here is similar to the cavity-mediated approach, but using a qubit instead of a waveguide to mediate the coupling. The difference with the second approach we presented is that in this case the three qubits interacts capacitively, and are not physically coupled.  In order to compute the effective ZZ interaction between the qubits, one has to perform two Schrieffer-Wolff transformations. The first transformation is needed to eliminate the degrees of freedom of the coupler and it is allowed when the qubits are detuned with respect to it. The second one can be performed when the two qubits are detuned with respect to each other, and it is needed to eliminate the SWAP term. The effective ZZ coupling was calculated up to fourth order in Refs.~\cite{Yan2018,Sung2021}. The first non-zero contribution for the coupling strength is approximately given by
\begin{equation}
    \zeta \approx - \frac{g^2(\alpha_A + \alpha_B)}{(\Delta_{AB}  +\alpha_A)(\Delta_{AB} - \alpha_B)}\,,
\end{equation}
with $\Delta_{AB} = \omega_A - \omega_B$ is the detuning between the two qubits, $\alpha_{\chi}$ is the anharmonicity of the $\chi$-type qubit and $g$ is the capacitive coupling between the qubits and the coupler.
This scheme was investigated experimentally in Ref.~\cite{Collodo2020}, where they reached a coupling strength of $\sim 50$ MHz, and numerically in Ref.~\cite{Baker2022}. It is worth mentioning that this type of building block is the one employed in modern quantum architectures, e.g.~in the Sycamore quantum processor~\cite{Arute2019}.

\section{Dynamical evolution of the 2D ladder}\label{sezDYN} 
In this Section we discuss  in  details the dynamical evolution of the 2D ladder.
The material is organized as follows: in Sec.~\ref{later} we solve the equation of motion for arbitrary sequences of
control pulses under the $e$--$e$ blockade regime proving that they allows us to realize 
a certain product of the operators $\hat{W}_{\chi}(\theta',\bm{n}'; \theta'',\bm{n}'')$ defined in Eq.~(6) of the main text; in Sec.~\ref{sec:gen} we present some generic
properties of  control unitaries $\hat{W}_{\chi}(\theta',\bm{n}'; \theta'',\bm{n}'')$; in Sec.~\ref{secL=:universal} we finally show that any possible 
transformations $\hat{W}_{\chi}(\theta',\bm{n}'; \theta'',\bm{n}'')$ can be implemented by  properly selecting
the sequences of our global  controls. 

\subsection{Effective dynamics under  $e$--$e$ blockade conditions}\label{later} 
As detailed in the main text, in our model the classical control functions $\Omega_\chi(t)$ and $\phi_\chi(t)$ associated with the pulses $V_A(t)$, $V_B(t)$,  and $V_C(t)$
assume constant values on disjoint time intervals. More precisely, in our analysis we assume that if $V_A(t)$ is active (not null), than both
$V_B(t)$ and $V_C(t)$ must be null, and vice-versa that if either $V_B(t)$ or $V_C(t)$ or both are active, than 
$V_A(t)$ must be null, i.e. 
\begin{eqnarray} \label{constraintDISJ} 
\left\{\begin{array}{l} 
V_A(t) \neq 0 \Longrightarrow V_B(t)=V_C(t)=0\;, 
\\
V_B(t) \neq 0  \Longrightarrow V_A(t)=0\;,  \\ 
V_C(t) \neq 0  \Longrightarrow V_A(t)=0\;,  
\end{array} \right.
\end{eqnarray} 
(notice that $V_B(t)$ and $V_C(t)$ are allowed to act on the system at the same time). Formally this implies that we can divide the temporal evolution
into a collection of disjoint time windows ${\cal T}_1, {\cal T}_2, \cdots, {\cal T}_\ell$ 
where the Hamiltonian $\hat{H}_{\text{rf}}(t)$ of Eq.~(4) of the main text can be treated as time-independent, i.e. 
\begin{eqnarray}
\hat{H}_{\text{rf}}(t) &=& \hat{H}^{(\ell)}_{\text{rf}}:= \sum_{\chi\in {X}_\ell}  
\sum_{i \in{\chi}} \frac{\hbar \Omega^{(\ell)}_{\chi}}{2}  \Big(e^{i\phi^{(\ell)}_{\chi}}\vert g_i \rangle \langle e_i \vert + {\rm H.c.} \Big)  + \sum_{\langle i,j \rangle} 2 \hbar \zeta \vert e_i e_{j} \rangle \langle e_i e_{j} \vert \;,   \qquad \forall t\in {\cal T}_\ell\;, 
\label{HfinalSUP} 
\end{eqnarray}
where ${X}_\ell$ is the subset of $\{ A, B,C\}$ that specifies which controls are active in the $\ell$-th time interval, 
and where, given $\chi \in X_\ell$, 
$\Omega^{(\ell)}_{\chi}$ and $\phi^{(\ell)}_{\chi}$ are the constant values assumed by the functions 
$\Omega_{\chi}(t)$ and $\phi_{\chi}(t)$ on such interval, respectively.
Notice that from (\ref{constraintDISJ}) it follows that the only allowed possibility for ${X}_\ell$ 
are $\{A\}$ ($A$-type qubit controlled, $B$- and $C$-type not), $\{ B\}$ ($B$-type qubit controlled, $A$- and $C$-type qubit not), $\{ C\}$ ($C$-type qubit controlled, $A$- and $B$-type qubit not), and finally 
$\{ B, C\}$ ($B$- and $C$-type qubit controlled, $A$-type qubit not).
Therefore the total evolution induced by  $\hat{H}_{\text{rf}}(t)$ can be expressed as the sequence in Eq.~(5) of the main text, i.e. 
\begin{eqnarray} \label{stringSUP} \hat{U}_{\rm tot} : =\overleftarrow{\exp}\left[ -\frac{i}{\hbar} \int_{{\cal T}_{\rm tot}} dt' \hat{H}_{\text{rf}}(t')\right] = 
\hat{U}^{(\ell)}\cdots \hat{U}^{(2)}\hat{U}^{(1)}\;, \end{eqnarray}
where $\overleftarrow{\exp}[ ... ]$ stands for the time-ordered exponential, ${\cal T}_{\rm tot}:= {\cal T}_1 \bigcup{\cal T}_2
\bigcup
 \cdots \bigcup {\cal T}_\ell$ is the total time interval of the dynamics, and 
 \begin{eqnarray} \label{UELL} 
 \hat{U}^{(\ell)} := \exp\left[-\frac{i}{\hbar}\hat{H}^{(\ell)}_{\text{rf}} \tau_\ell \right]\;, 
 \end{eqnarray} 
is the unitary evolution associated with the $\ell$-th time interval ($\tau_\ell$ being its  temporal duration). 
In the $e$--$e$ blockade regime where the coupling constant $\zeta$ is the main energy scale of the model (i.e. 
$\eta_{\rm BR} := |\zeta/\Omega^{(\ell)}_{\chi}|\gg 1$),
the transitions induced by the driving part of $\hat{H}^{(\ell)}_{\text{rf}}$ are suppressed whenever they involve
input or final states where at least one of the interacting neighbours of the $\chi$-type qubits affected by the control is in the excited state. More precisely we shall see  that in the limit $\eta_{\rm BR} \gg 1$ each element $\hat{U}^{(\ell)}$ of the sequence~(\ref{stringSUP}) can be expressed in terms of control unitary gates of the form 
\begin{eqnarray} 
\hat{U}^{(\ell)}_{\chi} :=  \prod_{i\in\chi} \left[  \hat{\openone}_i \otimes \hat{Q}_{\langle i \rangle}+\hat{\mathbb{U}}^{(\ell)}_i \otimes \hat{P}_{\langle i \rangle}  \right] \label{ccuSUP}
\;, 
\end{eqnarray} 
up to a global unitary gate that can be postponed to the
very end of the process and which plays no role in the process. In Eq.~(\ref{ccuSUP}) $\hat{P}_{\langle i \rangle}$ is the projector on the subspace of the nearest-neighbouring qubits of the $i$-th qubit which contains no excitations, $\hat{Q}_{\langle i \rangle}$ is  the orthogonal complement of $\hat{P}_{\langle i \rangle}$, while finally $\hat{\mathbb{U}}^{(\ell)}_i$ is the single-qubit unitary evolution induced by the Rabi contribution of the Hamiltonian $\hat{H}^{(\ell)}_{\text{rf}}$ on the $\chi$-type qubits, i.e.
\begin{eqnarray} \label{defuellsup} 
\hat{\mathbb{U}}^{(\ell)}_i &:=&\exp[ -i \frac{\Omega^{(\ell)}_{\chi}\tau_\ell}{2} 
\big(e^{i\phi^{(\ell)}_{\chi}}\vert g_i \rangle \langle e_i \vert + {\rm H.c.} \big)] = 
\exp[ -i \frac{\Omega^{(\ell)}_{\chi}\tau_\ell}{2}  \left(  \cos(\phi^{(\ell)}_{\chi})~\hat{\sigma}^{(x)}_i + \sin (\phi^{(\ell)}_{\chi})~\hat{\sigma}^{(y)}_i\right)]\;. 
\end{eqnarray} 
We recall that if the index  $i$ identifies a regular (possibly crossed) $\chi$-type qubit we have  
$\hat{P}_{\langle i \rangle}:= |g g \rangle\langle g g|$ and $\hat{Q}_{\langle i \rangle}:= |ee \rangle\langle ee| +
|eg \rangle\langle eg| + |ge \rangle\langle ge|$, where  $|g g \rangle$, $|e g \rangle$,
$|g e \rangle$, $|ee \rangle$, represent the energy levels of the two qubits  that exhibit a ZZ coupling with such qubit (notice that in case the $i$-th qubit is of $B$- or $C$- type, then such qubits will be always of $A$-type; while if the $i$-th qubit is of $A$-type, then they can be of $B$- or $C$-type). On the contrary if $i$ identifies a $B$ or a $C$ qubit  with a black circle then, the qubit is coupled with only a single $A$-type qubit so that 
$\hat{P}_{\langle i \rangle}:= |g  \rangle\langle g |$ and $\hat{Q}_{\langle i \rangle}:= |e \rangle\langle e|$.
Finally if instead  $i$ identifies a $B$-type or a $C$-type qubit with a black triangle then there are three interacting $A$-type qubits so that 
$\hat{P}_{\langle i \rangle}:= |g g g\rangle\langle g g g|$ and $\hat{Q}_{\langle i \rangle}:= |eee \rangle\langle eee| +
|eeg \rangle\langle eeg| + \cdots + |gge \rangle\langle gge|$.
\\
To prove Eq.~(\ref{ccuSUP}) it is convenient to 
express the unitary~(\ref{UELL})  in the interaction picture, identifying the interaction part  $\hat{H}_{\rm int}$ of the  Hamiltonian~(\ref{HfinalSUP})  with the term that  is proportional to the Rabi frequency, and the free-contribution $\hat{H}_{\rm free}$ with the term that is instead proportional to the ZZ coupling, i.e. 
\begin{eqnarray}
\hat{H}^{(\ell)}_{\text{rf}}&=& \hat{H}_{\rm free} + \hat{H}^{(\ell)}_{\rm int}\;, \qquad \left\{ 
\begin{array}{l} 
\hat{H}_{\rm free}:= \sum_{\langle i,j \rangle} 2 \hbar \zeta \vert e_i e_{j} \rangle \langle e_i e_{j} \vert \;, \label{newHFREE} \\\\ 
\hat{H}^{(\ell)}_{\rm int}:=\sum_{\chi\in {X}_\ell}   \sum_{i \in{\chi}} \frac{\hbar \Omega^{(\ell)}_{\chi}}{2}  \Big(e^{i\phi^{(\ell)}_{\chi}}\vert g_i \rangle \langle e_i \vert + {\rm H.c.} \Big)  \;.\label{HfinalSUP1} \end{array} \right.
\end{eqnarray} 
Accordingly we can then write 
\begin{eqnarray} \label{UELL1} 
\hat{U}^{(\ell)}&=& \exp\left[-\frac{i}{\hbar}  \hat{H}^{(\ell)}_{\text{rf}} \tau_{\ell}\right] = 
\exp\left[-\frac{i}{\hbar}  \hat{H}_{\rm free} \tau_{\ell}\right]\overleftarrow{\exp}\left[ -\frac{i}{\hbar} \int_{{\cal T}_{\ell}} dt' \hat{H}^{\prime(\ell)}_{\rm int}(t')\right] \;, \end{eqnarray} 
where $\tau_{\ell}$ is the duration of the time  window  ${\cal T}_{\ell}$ and  
\begin{eqnarray} 
\hat{H}^{\prime(\ell)}_{\rm int}(t)&:=& \exp\left[\frac{i}{\hbar}  \hat{H}_{\rm free} t\right]
\hat{H}^{(\ell)}_{\rm int} \exp\left[-\frac{i}{\hbar}  \hat{H}_{\rm free}t \right]\;, 
\end{eqnarray} 
the interaction term expressed in the interaction picture. 
To determine $\hat{H}^{\prime(\ell)}_{\rm int}(t)$ observe that the various contributions of $\hat{H}_{\rm free}$ all commute so that we can write  
\begin{eqnarray} \label{eHfree} 
e^{-\frac{i}{\hbar}  \hat{H}_{\rm free} t} = \prod_{\langle i,j \rangle} e^{-{i} \zeta \vert e_i e_{j} \rangle \langle e_i e_{j} \vert t} =  \prod_{\langle i,j \rangle} \left( \hat{\openone}_{ij} + (e^{-{i} \zeta  t}-1) \vert e_i e_{j} \rangle \langle e_i e_{j} \vert \right)\;, 
\end{eqnarray} 
where $\hat{\openone}_{ij}$ is the identity operator on the Hilbert space of the $i$-th and $j$-th qubit
that form an interacting ZZ couple $\langle i, j\rangle$. Observe also that  when acting on a  generic product state $|\vec{J}\rangle$ of the ladder qubits formed by elements of the computational basis  $\{ |g\rangle, |e\rangle\}$ (e.g. states of the form $| e_1 g_1 e_2e_3 \cdots g_M \rangle$ with $M$ being the total number of qubit in the system), the transformation $e^{-\frac{i}{\hbar}  \hat{H}_{\rm free} t}$ will evolve it by adding a phase contribution $e^{-i\zeta t}$ for each of the ZZ interacting couples which are in the $|ee\rangle$ configuration. 
More explicitly we  can write 
\begin{eqnarray} \label{evolutionfreee} 
e^{-\frac{i}{\hbar}  \hat{H}_{\rm free} t} |\vec{J} \rangle = e^{-{i} M({\vec{J}})  \zeta  t}| \vec{J} \rangle\;, 
\end{eqnarray} 
with $M(\vec{J})$ the integer that counts the number of ZZ interacting couples which in the sequence $|\vec{J}\rangle$ are  in the $|ee\rangle$ state. To see how $e^{-\frac{i}{\hbar}  \hat{H}_{\rm free} t}$ transforms $\hat{H}_{\rm int}$ it is useful to recall that the  $|g_i\rangle\langle e_i|$ which enters in Eq.~(\ref{HfinalSUP1}) is an operator on the full ladder, that for all qubits but the $i$-th, acts as the identity. In particular, indicating with $|\vec{J}^{(i)} \rangle$ a generic element of the computational
basis $\{ |g\rangle, |e\rangle\}$ of the entire collection of the  qubits of the system excluded the $i$-th we can write 
\begin{eqnarray} 
|g_i\rangle\langle e_i| \equiv \sum_{\vec{J}^{(i)}}  |g_i \vec{J}^{(i)} \rangle\langle e_i\vec{J}^{(i)} | \;.\label{deco} 
\end{eqnarray} 
From Eq.~(\ref{evolutionfreee}) it hence follows that 
\begin{eqnarray} 
e^{\frac{i}{\hbar}  \hat{H}_{\rm free} t}|g_i\rangle\langle e_i|e^{-\frac{i}{\hbar}  \hat{H}_{\rm free} t } =
\sum_{\vec{J}^{(i)}}  e^{-{i} [ M_g(\vec{J}^{(i)})-M_e(\vec{J}^{(i)})] \zeta t} |g_i \vec{J}^{(i)} \rangle\langle e_i\vec{J}^{(i)} |\;,
\label{deco11} 
\end{eqnarray} 
with $M_g(\vec{J}^{(i)})$ and $M_e(\vec{J}^{(i)})$ counting respectively the number of ZZ interacting couples which in the sequences $|g_i \vec{J}^{(i)} \rangle$ and   $|e_i \vec{J}^{(i)} \rangle$
are  in the $|ee\rangle$ state. 
Observe that by construction we have that 
$M_e(\vec{J}^{(i)})$ is always greater than or equal to $M_g(\vec{J}^{(i)})$, and that the two coincide if and only if the qubits with which the $i$-th qubits are ZZ-coupled, 
are all in the ground state, a condition which can expressed as  
\begin{eqnarray} 
\left\{ \begin{array}{l} 
\hat{Q}_{\langle i \rangle} |\vec{J}^{(i)} \rangle = |\vec{J}^{(i)} \rangle  \quad 
\Longrightarrow \quad M_e(\vec{J}^{(i)}) \geq    M_g(\vec{J}^{(i)})  + 1 \;,  \\ \\ 
\hat{P}_{\langle i \rangle} |\vec{J}^{(i)} \rangle = |\vec{J}^{(i)} \rangle  \quad 
\Longrightarrow \quad M_e(\vec{J}^{(i)})=  M_g(\vec{J}^{(i)}) \;,
\end{array} \right. 
\end{eqnarray} 
with $\hat{P}_{\langle i \rangle}$ and  $\hat{Q}_{\langle i \rangle}$ the projectors appearing in Eq.~(\ref{ccuSUP}). 
Accordingly we can write 
\begin{eqnarray} 
e^{\frac{i}{\hbar}  \hat{H}_{\rm free} t}|g_i\rangle\langle e_i|e^{-\frac{i}{\hbar}  \hat{H}_{\rm free} t } &=&
\sum_{\vec{J}^{(i)}: \hat{P}_{\langle i \rangle} |\vec{J}^{(i)} \rangle = |\vec{J}^{(i)} \rangle}   |g_i \vec{J}^{(i)} \rangle\langle e_i\vec{J}^{(i)} |
+\sum_{\vec{J}^{(i)}: \hat{Q}_{\langle i \rangle} |\vec{J}^{(i)} \rangle = |\vec{J}^{(i)} \rangle}   e^{-{i} [ M_g(\vec{J}^{(i)})-M_e(\vec{J}^{(i)})] \zeta t} |g_i \vec{J}^{(i)} \rangle\langle e_i\vec{J}^{(i)} |\nonumber \\
&=&    |g_i \rangle\langle e_i |\otimes \hat{P}_{\langle i \rangle} + \hat{\Delta}_{i}(t) \;, 
\label{deco111} 
\end{eqnarray} 
with  
\begin{eqnarray} \hat{\Delta}_{i}(t):= \sum_{\vec{J}^{(i)}: \hat{Q}_{\langle i \rangle} |\vec{J}^{(i)} \rangle = |\vec{J}^{(i)} \rangle}   e^{-{i} [ M_g(\vec{J}^{(i)})-M_e(\vec{J}^{(i)})] \zeta t} |g_i \vec{J}^{(i)} \rangle\langle e_i\vec{J}^{(i)} |=
\hat{Q}_{\langle i \rangle} e^{\frac{i}{\hbar}  \hat{H}_{\rm free} t}|g_i\rangle\langle e_i|e^{-\frac{i}{\hbar}  \hat{H}_{\rm free} t }  \hat{Q}_{\langle i \rangle}
\;, 
\end{eqnarray} 
an operator made of a sum of terms which  oscillate with frequencies which are  integer multiples of $\zeta$. At the level of the operator
$\hat{H}^{\prime(\ell)}_{\rm int}(t)$ this translates into the identity 
\begin{eqnarray} 
\hat{H}^{\prime(\ell)}_{\rm int}(t)&=&\sum_{\chi\in {X}_\ell}\left(
\sum_{i\in \chi} \hat{H}^{(\ell)}_{\chi,i} \otimes \hat{P}_{\langle i \rangle} +\sum_{i\in \chi}   \hat{\Delta}_i(t) \right)\;, 
\\ 
\hat{H}^{(\ell)}_{\chi,i}&:=& \frac{\hbar \Omega^{(\ell)}_{\chi}}{2}  \Big(e^{i\phi^{(\ell)}_{\chi}}\vert g_i \rangle \langle e_i \vert + {\rm H.c.} \Big)\;.
\end{eqnarray} 
Under the $e$--$e$ blockade regime, $\eta_{\rm BR} = |\zeta/\Omega^{(\ell)}_{\chi_{\ell}}|\gg 1$, the oscillatory part of  $\hat{H}^{\prime(\ell)}_{\rm int}(t)$ evolves over time-scales that are much smaller than the typical
time-scales determined by the first contribution. 
Enforcing a temporal coarse-graining over time
intervals $T_{\rm c.g.}$ such that $1/|\Omega^{(\ell)}_{\chi_{\ell}}| \gg T_{\rm c.g.} \gg 1/|\zeta|$ this leads to 
\begin{eqnarray} \label{imo11}  \hat{H}^{\prime(\ell)}_{\rm int}(t)\Big|_{\eta_{\rm BR}\gg 1}  \simeq 
\int_{t}^{t+T_{\rm c.r.}} \frac{dt'}{T_{\rm c.r.}}    \hat{H}^{\prime}_{\rm int}(t')= 
\sum_{\chi\in {X}_\ell}  \sum_{i\in \chi} \hat{H}^{(\ell)}_{\chi,i} \otimes \hat{P}_{\langle i \rangle} \;.
\end{eqnarray} 
Accordingly we can write 
\begin{eqnarray} 
\overleftarrow{\exp}\left[ -\frac{i}{\hbar} \int_{{\cal T}_{\rm tot}} dt' \hat{H}^{\prime}_{\rm int}(t')\right] \Big|_{\eta_{\rm BR}\gg 1}
&\simeq&  \exp[ -\frac{i}{\hbar} \sum_{\chi\in {X}_\ell}  \sum_{i\in \chi} \hat{H}^{(\ell)}_{\chi,i} \otimes \hat{P}_{\langle i \rangle} \tau_\ell] 
=  \prod_{\chi\in X_\ell}
{\exp}\left[ -\frac{i}{\hbar}    \sum_{i\in \chi} \hat{H}^{(\ell)}_{\chi,i} \otimes \hat{P}_{\langle i \rangle}  \tau_\ell \right] \nonumber\\
&=& 
\prod_{\chi\in X_\ell} \prod_{i\in\chi} \left[  \hat{\openone}_i \otimes \hat{Q}_{\langle i \rangle}+\hat{\mathbb{U}}^{(\ell)}_i \otimes \hat{P}_{\langle i \rangle}  \right]
\label{UELL11} 
\;, \end{eqnarray} 
with $\hat{\mathbb{U}}_i^{(\ell)}$ as in Eq.~(\ref{defuellsup}). Notice that in writing the last two identities we do not need to worry about the
ordering of the various terms since they all commute. In particular the commutation of the terms associated with the different species of the subset $X_\ell$ is a direct consequence  of the  constraints~(\ref{constraintDISJ})  which prevents the possibility of driving two interacting species (i.e. $AB$ or $AC$) in the same time window. 
Replacing this into Eq.~(\ref{UELL1}) we hence obtain that for each time window ${\cal T}_\ell$ one has
\begin{eqnarray} \label{UELL111} 
\hat{U}^{(\ell)}\Big|_{\eta_{\rm BR}\gg 1}\simeq 
e^{-\frac{i}{\hbar}  \hat{H}_{\rm free} \tau_\ell}\; 
\left(  \prod_{\chi\in X_\ell} \prod_{i\in\chi} \left[  \hat{\openone}_i \otimes \hat{Q}_{\langle i \rangle}+\hat{\mathbb{U}}^{(\ell)}_i \otimes \hat{P}_{\langle i \rangle}  \right]\right)  = e^{-\frac{i}{\hbar}  \hat{H}_{\rm free} \tau_\ell}\; 
\prod_{\chi\in X_\ell} \hat{U}^{(\ell)}_{\chi} 
\;.\end{eqnarray} 
Noticing  that $e^{-\frac{i}{\hbar}  \hat{H}_{\rm free} \tau_\ell}$ commutes with~(\ref{imo11}) and hence with all the operators of the form (\ref{UELL11}) (see e.g. Eq.~(\ref{deco111})), this finally allows us to express 
the global evolution operator (\ref{stringSUP}) as
\begin{eqnarray} \label{SstringSUP} \hat{U}_{\rm tot} \Big|_{\eta_{\rm BR}\gg 1} &\simeq &
e^{-\frac{i}{\hbar}  \hat{H}_{\rm free} \tau_{\rm tot}}\; \prod_{\chi_\ell \in X_\ell}  \cdots \prod_{\chi_2\in X_2}
\prod_{\chi_1\in X_1}
\hat{U}^{(\ell)}_{\chi_\ell} \cdots \hat{U}^{(2)}_{\chi_2} \; \hat{U}^{(1)}_{\chi_1}
\;, \end{eqnarray} 
proving the thesis.

\subsubsection*{Dynamics of a $ABACA$ qubit chain}
In this section, we discuss the generalized case of driving a chain composed of $ABACA$ qubits. Recall that each of the three qubit species can be controlled independently at different time intervals: specifically, $\Omega_A(t) \neq 0 \Rightarrow \Omega_B(t) = \Omega_C(t) = 0$, and $ \Omega_B(t) \neq 0$ or $\Omega_C(t) \neq 0 \Rightarrow \Omega_A(t) = 0$. However, as seen in the transfer protocol shown in Fig.~2(b) of the main text, the $B$- and $C$-type qubits are driven simultaneously, allowing $B$-type qubits to function as $C$-type qubits and vice versa. Here we emphasize that the inclusion of a third species, $C$, serves only to reduce the number of physical qubits required—specifically by half, compared to a two-species architecture. Moreover, we stress that we need to separately drive the $B$ and $C$ qubits only when single qubit gates are performed.
Referring to Fig.~2(a) of the main text, we consider $ ABA $ as the fundamental building block of our ladder. The $ABACA$ chain is then constructed by combining two building blocks: $ABA$ and $ACA$. If we drive the $B$- and $C$-type qubits with frequencies $\omega_{\mathrm{d},B} = \omega_{B} - 2\zeta$ and $\omega_{\mathrm{d},C} = \omega_{C} - 2\zeta$, respectively, we expect to enter a blockade regime where only transitions between the $\vert ggggg \rangle$ and $\vert gegeg \rangle$ states are allowed. The energy difference $\Delta E$ between these states can be explicitly calculated as:
\begin{equation}
\Delta E := E_{\vert ggggg \rangle} - E_{\vert gegeg \rangle} = (\omega_{B} - 2\zeta) + (\omega_{C} - 2\zeta),
\end{equation}
confirming the result shown in Fig.~2(a) of the main text. This $\Delta E$ represents the gap between these states, effectively selecting only these transitions and thus emulating the blockade mechanism along an arbitrarily long chain.

\subsection{General properties} \label{sec:gen} 
A convenient way to rewrite the operators $\hat{U}^{(\ell)}_{\chi}$   defined in Eq.~(\ref{ccuSUP}) is to introduce the parameters 
\begin{eqnarray} 
\left\{ \begin{array}{l} \theta^{(\ell)} : = \Omega^{(\ell)}_{\chi}\tau_\ell\;, \\ \\
\bm{n}_\perp^{(\ell)}:= \left(\cos(\phi^{(\ell)}_{\chi}), \sin (\phi^{(\ell)}_{\chi}),0\right)\;. \end{array} \right.
\end{eqnarray} 
Recalling that  the crossed elements of 
the $\chi$-type qubits have twice the  Rabi frequency of the regular ones, we can hence write   
\begin{eqnarray} 
\hat{U}^{(\ell)}_{\chi}
&=&  
\prod_{i\in\chi_{\ell}^{\rm r}} \left[  \hat{\openone}_i \otimes \hat{Q}_{\langle i \rangle}+\hat{\mathbb{R}}_i(\theta^{(\ell)},\bm{n}_\perp^{(\ell)}) \otimes \hat{P}_{\langle i \rangle}  \right]  \prod_{i\in\chi_{\ell}^{\times}} \left[  \hat{\openone}_i \otimes \hat{Q}_{\langle i \rangle}+\hat{\mathbb{R}}_i(2\theta^{(\ell)},\bm{n}_\perp^{(\ell)}) \otimes \hat{P}_{\langle i \rangle}  \right]  \nonumber \\\label{ubetter} 
&=&  \hat{W}_{\chi}(\theta^{(\ell)}, \bm{n}_\perp^{(\ell)}; 2\theta^{(\ell)}, \bm{n}_\perp^{(\ell)}) \;, 
\end{eqnarray} 
where defining $\chi^{\rm r}$ and $\chi^{\times}$ as the regular (non crossed) and crossed
subsets of $\chi$-type qubits, we introduced 
\begin{eqnarray} \label{WTRANSF}\left\{  \begin{array}{l} 
\hat{W}_{\chi}(\theta',\bm{n}'; \theta'',\bm{n}'') :=
\hat{W}_{\chi^{\rm r}}(\theta',\bm{n}')  \; \hat{W}_{\chi^{\times}}(\theta'',\bm{n}'')\;, \\\\
\hat{W}_{\xi}(\theta,\bm{n}):= 
\prod_{i\in \xi} \left[  \hat{\openone}_i \otimes \hat{Q}_{\langle i \rangle}+
\hat{\mathbb{R}}_i(\theta,\bm{n}) \otimes \hat{P}_{\langle i \rangle}  \right]\;,  \qquad \xi\in \{ \chi^{\rm r},\chi^{\times}\}\end{array} \right.
\end{eqnarray} 
with 
\begin{eqnarray} 
\hat{\mathbb{R}}_i(\theta,\bm{n})
     &:=&\exp[ -i ({\theta}/{2}) \bm{n} \cdot \vec{\sigma}_i]=\cos(\theta/2) \hat{\openone}_i
    - i \sin(\theta/2) \bm{n} \cdot \vec{\sigma}_i\;, 
      \qquad \qquad 
     \vec{\sigma}_i:=(\hat{\sigma}^{(x)}_i,\hat{\sigma}^{(y)}_i,\hat{\sigma}^{(z)}_i)\;, 
 \end{eqnarray} 
the single-qubit unitary rotation associated with the unit vector $\bm{n}$ and the angle $\theta$.
 \\ 
 
The transformations~(\ref{WTRANSF}) obey some very useful properties:
\begin{enumerate}
\item The operators $\hat{W}_{B}(\theta_B',\bm{n}_B'; \theta_B'',\bm{n}''_B)$ and\
$\hat{W}_{C}(\theta_C',\bm{n}_C'; \theta_C'',\bm{n}_C'')$ always commute, i.e. 
\begin{eqnarray}  \hat{W}_{B}(\theta_B',\bm{n}_B'; \theta_B'',\bm{n}''_B)  \hat{W}_{C}(\theta_C',\bm{n}_C'; \theta_C'',\bm{n}_C'')= \label{COMMBC} 
 \hat{W}_{C}(\theta_C',\bm{n}_C'; \theta_C'',\bm{n}_C'')\hat{W}_{B}(\theta_B',\bm{n}_B'; \theta_B'',\bm{n}''_B) 
\;.
   \end{eqnarray}
This is due to the fact that both the $B$ and the $C$ elements are not directly coupled by the ZZ interaction (see Fig.~1 of the main text). 
They only share ZZ coupling  with the $A$-type qubits. On the contrary neither $\hat{W}_{B}(\theta_B',\bm{n}_B'; \theta_B'',\bm{n}''_B)$ nor $\hat{W}_{C}(\theta_C',\bm{n}_C'; \theta_C'',\bm{n}_C'')$ in general commute with
$\hat{W}_{A}(\theta_A',\bm{n}_A'; \theta_A'',\bm{n}_A'')$.

\item For fixed $\chi\in \{A,B,C\}$ setting $\theta'=0$ ($\theta''=0$), the unitary evolution $\hat{W}_{\chi}(\theta',\bm{n}'; \theta'',\bm{n}'')$
reduces to the mapping $\hat{W}_{\chi^{\times}}(\theta'',\bm{n}'')$ ($\hat{W}_{\chi^{\rm r}}(\theta',\bm{n}')$)
that only act on the crossed (regular) elements of the $\chi$-type qubits, i.e. 
\begin{eqnarray}  \hat{W}_{\chi^{\rm r}}(\theta',\bm{n}')  =\hat{W}_{\chi}(\theta',\bm{n}'; 0,\bm{n}'') \;, \qquad 
 \hat{W}_{\chi^{\times}}(\theta'',\bm{n}'')  =\hat{W}_{\chi}(0,\bm{n}'; \theta'',\bm{n}'')\;,
   \end{eqnarray}
   (this is a trivial consequence of the fact that $\hat{P}_{\langle i \rangle}+ \hat{Q}_{\langle i \rangle}=\hat{\openone}_{\langle i\rangle}$ is the identity operator on the associated space).

\item For fixed $\chi\in \{A,B,C\}$, the crossed and regular contributions of $\hat{W}_{\chi}(\theta',\bm{n}'; \theta'',\bm{n}'')$ commute, i.e. 
\begin{eqnarray}  \hat{W}_{\chi^{\rm r}}(\theta',\bm{n}')   \hat{W}_{\chi^{\times}}(\theta'',\bm{n}'')=
  \hat{W}_{\chi^{\times}}(\theta'',\bm{n}'') \hat{W}_{\chi^{\rm r}}(\theta',\bm{n}') \;.
  \end{eqnarray}
   \item For fixed $\chi\in \{A,B,C\}$ and $\xi\in \{ \chi^{\rm r},\chi^{\times}\}$  the transformations 
  $\hat{W}_{\xi}(\theta,\bm{n})$ form a group that  obeys the same composition rules of the $SU(2)$ 
  defined by unitary matrices $\hat{\mathbb{R}}_i(\theta,\bm{n})$.
  In particular we have  that, given the angles  $\theta_1,\theta_2$ and the unit vectors 
  $\bm{n}_1$, $\bm{n}_2$, one has 
   \begin{eqnarray}  
   \hat{W}_{\xi}(\theta_2,\bm{n}_2)   \hat{W}_{\xi} (\theta_1,\bm{n}_1)=
   \hat{W}_{\xi}(\theta_3,\bm{n}_3)\;, \label{compo}
  \end{eqnarray}
  with the angle $\theta_3$ and the unit vector $\bm{n}_3$ satisfying the identity 
  $\hat{\mathbb{R}}_i(\theta_2,\bm{n}_2) \hat{\mathbb{R}}_i(\theta_1,\bm{n}_1) = \hat{\mathbb{R}}_i(\theta_3,\bm{n}_3)$.
  Furthermore  
   the inverse of $\hat{W}_{\xi}(\theta,\bm{n})$ corresponds to 
   $\hat{W}_{\xi}(-\theta,\bm{n})= \hat{W}_{\xi}(\theta,-\bm{n})$, i.e. 
   \begin{eqnarray} 
   \hat{W}^{-1}_{\xi}(\theta,\bm{n})=\hat{W}^{\dag}_{\xi}(\theta,\bm{n})= \hat{W}_{\xi}(-\theta,\bm{n})= \hat{W}_{\xi}(\theta,-\bm{n})\;. 
   \end{eqnarray} 
   \item From the algebra of  the single qubit rotations $\hat{\mathbb{R}}_i(\theta,\bm{n})$, it follows that   for fixed $\chi\in \{A,B,C\}$ and $\xi\in \{ \chi^{\rm r},\chi^{\times}\}$, given $\bm{n}$ an arbitrary unit vector, the transformations $\hat{W}_{\xi}(\theta,\bm{n})$ are periodic in $\theta$ with period $4\pi$, i.e.
     \begin{eqnarray}  
   \hat{W}_{\xi}(4\pi+\theta,\bm{n})  =  \hat{W}_{\xi} (\theta,\bm{n})\;.\end{eqnarray}
   In particular for $\theta=4\pi$ the transformation coincides with the identity. Notice however that for $\theta=2\pi$
   the evolution corresponds to a non trivial phase-transformation on the $\xi$-type qubits,
   \begin{eqnarray} 
  \left\{ \begin{array}{l} \hat{W}_{\xi}(4\pi,\bm{n})  =  \hat{W}_{\xi} (0,\bm{n})= 
   \prod_{i\in\xi}  \hat{\openone}_i\otimes  \left[  \hat{Q}_{\langle i \rangle} +   \hat{P}_{\langle i \rangle} \right]=\hat{\openone} 
   \;, \\  \\
      \hat{W}_{\xi}(2\pi,\bm{n})  =    \hat{Z}_{\xi}^{(\rm tot)}:=   
   \prod_{i\in\xi}  \hat{\openone}_i\otimes  \left[  \hat{Q}_{\langle i \rangle} -   \hat{P}_{\langle i \rangle} \right]=
     \prod_{i\in\xi}   \left[  \hat{Q}_{\langle i \rangle} -   \hat{P}_{\langle i \rangle} \right]\;. \end{array} \right. \label{defZTOT} 
  \end{eqnarray}
  \item The property~(\ref{compo}) for $\chi^{\rm r}$ and $\chi^{\times}$ translates in the following composition rule
  for the global operations, i.e. 
  \begin{eqnarray} \label{WTRANSF1q} 
  \hat{W}_{\chi}(\theta^{\prime}_2,\bm{n}^{\prime}_2; \theta^{\prime\prime}_2,\bm{n}^{\prime\prime}_2)
 \hat{W}_{\chi}(\theta^{\prime}_1,\bm{n}^{\prime}_1; \theta^{\prime\prime}_1,\bm{n}^{\prime\prime}_1)&=&
 \hat{W}_{\chi}(\theta^{\prime}_3,\bm{n}^{\prime}_3; \theta^{\prime\prime}_3,\bm{n}^{\prime\prime}_3)\;,
 \end{eqnarray} 
with $\theta^{\prime}_3$, $\bm{n}^{\prime}_3$, and $\theta^{\prime\prime}_4$, $\bm{n}^{\prime\prime}_4$ 
determined by
the identities $\hat{\mathbb{R}}_i(\theta^{\prime}_2,\bm{n}^{\prime}_2) \hat{\mathbb{R}}_i(\theta^{\prime}_1,\bm{n}^{\prime}_1) = \hat{\mathbb{R}}_i(\theta^{\prime}_3,\bm{n}^{\prime}_3)$ and 
$\hat{\mathbb{R}}_i(\theta^{\prime\prime}_2,\bm{n}^{\prime\prime}_2) \hat{\mathbb{R}}_i(\theta^{\prime\prime}_1,\bm{n}^{\prime\prime}_1) = \hat{\mathbb{R}}_i(\theta^{\prime\prime}_3,\bm{n}^{\prime\prime}_3)$,
respectively.
\end{enumerate} 

\subsection{Universal control} \label{secL=:universal}
In Sec.~\ref{later} we have seen that in the $e$--$e$ blockade regime, each individual  pulsed unitary  $\hat{U}^{(\ell)}$ that compose $\hat{U}_{\rm tot}$ implements a specific type of control unitary gates~(\ref{WTRANSF}), i.e. the subset of transformations
\begin{eqnarray} \label{fdfsdf} 
\hat{W}_{\chi}(\theta, \bm{n}_\perp; 2\theta, \bm{n}_\perp) = \hat{W}_{\chi^{\rm r}}(\theta, \bm{n}_\perp)  \;  \hat{W}_{\chi^{\times}}(2\theta, \bm{n}_\perp)\;,
\end{eqnarray} 
which act simultaneously on $\chi^{\rm r}$ and $\chi^{\times}$ inducing single-qubit rotations around a
generic unit vector $\bm{n}_\perp$ of the $xy$-plane (i.e. $\bm{z}\cdot \bm{n}_\perp=0$) and with correlated angles $\theta$ and $2\theta$. Here we show that these constraints can be overcome by properly combining sequences of these special type of pulses: specifically we prove that the sequences~(\ref{stringSUP}) generated by the classical controls of the ladder, can produce any  transformation $\hat{W}_{\chi}(\theta',\bm{n}'; \theta''\bm{n}'')$ with $\theta'$, $\theta''$, and $\bm{n}'$, $\bm{n}''$ arbitrarily selected. The first ingredient to attain this result is  a property that holds for Pauli rotations along orthogonal unit vectors, i.e. 
\begin{eqnarray}
\bm{n}\cdot \bm{m}=0  \quad \Longrightarrow \quad  \hat{\mathbb{R}}_i(\pi,-\bm{n})\hat{\mathbb{R}}_i(\theta,\bm{m}) \hat{\mathbb{R}}_i(\pi,\bm{n})= \hat{\mathbb{R}}_i(-\theta,\bm{m}) =\hat{\mathbb{R}}^{-1}_i(\theta,\bm{m}) \qquad \forall \theta\;. 
\end{eqnarray} 
This identity is a direct consequence of the fact that $\hat{\mathbb{R}}_i(\pi,\bm{n})= -i \bm{n}\cdot \vec{\sigma}_i$, and that the Pauli operators $(\bm{n}\cdot \vec{\sigma}_i)$ and $(\bm{m}\cdot \vec{\sigma}_i)$ anti-commute, i.e.  
$(\bm{n}\cdot \vec{\sigma}_i)(\bm{m}\cdot \vec{\sigma}_i) (\bm{n}\cdot \vec{\sigma}_i)=-\bm{m}\cdot \vec{\sigma}_i$.
Recalling that $\hat{\mathbb{R}}_i(2\pi,\bm{n})=-\hat{\openone}_i$,   we can hence conclude that  
\begin{eqnarray}\label{impoide} 
\left\{ \begin{array}{l} \hat{\mathbb{R}}_i(\pi,-\bm{n})\hat{\mathbb{R}}_i(\theta,\bm{m}) \hat{\mathbb{R}}_i(\pi,\bm{n})
 \hat{\mathbb{R}}_i(\theta,\bm{m})
 = \hat{\mathbb{R}}_i(0,\bm{m})=\hat{\openone}_i \;, \\\\
 \hat{\mathbb{R}}_i(2\pi,-\bm{n})\hat{\mathbb{R}}_i(2\theta,\bm{m}) \hat{\mathbb{R}}_i(2\pi,\bm{n})
 \hat{\mathbb{R}}_i(2\theta,\bm{m})
 = \hat{\mathbb{R}}_i(4\theta,\bm{m})\;. 
\end{array} \right.
\end{eqnarray} 
Consider hence a 4-pulse sequence~(\ref{stringSUP}), $\hat{U}_{\chi}^{(4)}
\hat{U}_{\chi}^{(3)}
\hat{U}_{\chi}^{(2)}
\hat{U}_{\chi}^{(1)}$, where the first and the third element, $\hat{U}_{\chi}^{(1)}$ and  $\hat{U}_{\chi}^{(3)}$, correspond to the same transformation $\hat{W}_{\chi}(\theta/4, \bm{n}_\perp; \theta/2, \bm{n}_\perp)$ 
(see Eq.~(\ref{ubetter})) with assigned axis $\bm{n}_\perp$ in the $xy$-plane. On the contrary 
the second and fourth term, $\hat{U}_{\chi}^{(2)}$ and  $\hat{U}_{\chi}^{(4)}$, are equal to 
$\hat{W}_{\chi}(\pi, \bm{m}_\perp; 2\pi, \bm{m}_\perp)$ and 
$\hat{W}_{\chi}(\pi, -\bm{m}_\perp; 2\pi, -\bm{m}_\perp)$, respectively with
$\bm{m}_\perp$ also in the $xy$-plane but orthogonal to $\bm{n}_\perp$ (i.e.
$\bm{m}_\perp \cdot \bm{n}_\perp=0$).
Using Eq.~(\ref{SstringSUP}),  the composition rule given in Eq.~(\ref{WTRANSF1q}) and the identities~(\ref{impoide})
  we can hence conclude that
\begin{eqnarray}
 \hat{U}_{\chi}^{(4)}
\hat{U}_{\chi}^{(3)}
\hat{U}_{\chi}^{(2)}
\hat{U}_{\chi}^{(1)} \Big|_{\eta_{\rm BR}\gg 1} &\simeq&  \hat{W}_{\chi}(\pi, -\bm{m}_\perp; 2\pi, -\bm{m}_\perp)\hat{W}_{\chi}(\theta/4, \bm{n}_\perp; \theta/2, \bm{n}_\perp)
\hat{W}_{\chi}(\pi, \bm{m}_\perp; 2\pi, \bm{m}_\perp)\hat{W}_{\chi}(\theta/4, \bm{n}_\perp; \theta/2, \bm{n}_\perp) \nonumber \\ \label{firstfirst} 
&=& \hat{W}_{\chi}(0, \bm{n}_\perp; \theta, \bm{n}_\perp)= \hat{W}_{\chi^{\times}}(\theta, \bm{n}_\perp)\;,
\end{eqnarray} 
which for sake of simplicity we write dropping the contribution $\exp\left[-\frac{i}{\hbar}  \hat{H}_{\rm free} \tau_{\rm tot} \right]$. 
 Equation~(\ref{firstfirst}) implies that using the control pulses of the model we can 
 generate transformations that act selectively on the crossed elements of the $\chi$-type qubits inducing arbitrary
 rotations around any axis $\bm{n}_\perp$ in the $xy$-plane (in particular we can induce rotations around the
 $\bm{x}$ and $\bm{y}$ axis).
Invoking the standard Euler rotation theorem~\cite{Nielsen2010} we can generalize this to any control-unitary  $\hat{W}_{\chi^{\times}}(\theta, \bm{n})$ with respect to any possible (non necessarily orthogonal to $\bm{z}$) axis $\bm{n}$. Indeed recall that given $(\theta, \bm{n})$ arbitrary, there exist three angles $\alpha$, $\beta$, and $\gamma$, such that we can write 
  \begin{eqnarray}\label{euler} 
 \hat{\mathbb{R}}_i(\alpha,\bm{x})\hat{\mathbb{R}}_i(\beta,\bm{y}) \hat{\mathbb{R}}_i(\gamma,\bm{x})=
 \hat{\mathbb{R}}_i(\theta,\bm{n})
\;.
 \end{eqnarray}
Using sequences of the form~(\ref{firstfirst}) we can hence translate the above identity at the level of our control unitary
gates, obtaining 
  \begin{eqnarray}\label{euler1} 
 \hat{W}_{\chi^\times} (\alpha,\bm{x}) \hat{W}_{\chi^\times}(\beta,\bm{y})  \hat{W}_{\chi^\times}(\gamma,\bm{x})=
  \hat{W}_{\chi^\times}(\theta,\bm{n})
\;.
 \end{eqnarray}
Notice that to achieve this goal one needs no more than $3\times 4=12$ unitary pulses $\hat{U}_{\chi}$. 
A similar result can be applied to the control unitaries that act selectively on the regular elements of the $\chi$-type qubits. For instance using the transformation~(\ref{firstfirst}) we can compensate the $\chi^{\times}$ component of the gates~(\ref{fdfsdf}), i.e. 
 \begin{eqnarray} 
\hat{W}_{\chi^{\times}}(2\theta, -\bm{n}_\perp)
 \hat{W}_{\chi}(\theta, \bm{n}_\perp; 2\theta, \bm{n}_\perp) = 
  \hat{W}_{\chi^{\rm r}}(\theta, \bm{n}_\perp)  \;,
 \end{eqnarray} 
which shows that also for the regular  elements of the $\chi$-type qubits we can induce arbitrary
 rotation around any axis $\bm{n}_\perp$ in the $xy$-plane. 
From this we can then use the argument that led us to (\ref{euler1}) to conclude that sequences unitary pulses $\hat{U}_{\chi}$ can leads to any operations of the form  $\hat{W}_{\chi^{\rm r}}(\theta, \bm{n})$ with arbitrary
$\bm{n}$ and $\theta$, i.e. 
  \begin{eqnarray}\label{euler2} 
 \hat{W}_{\chi^{\rm r}} (\alpha,\bm{x}) \hat{W}_{\chi^{\rm r}}(\beta,\bm{y})  \hat{W}_{\chi^{\rm r}}(\gamma,\bm{x})=
  \hat{W}_{\chi^{\rm r}}(\theta,\bm{n})
\;.
 \end{eqnarray}
 Finally concatenating the results of Eqs.~(\ref{euler1}) and (\ref{euler2}) we can induce any arbitrary transformations of the form $\hat{W}_{\chi}(\theta',\bm{n}'; \theta'',\bm{n}'')$. 

\section{Information encoding and quantum computing}\label{secENC} 
\begin{figure}[t!]
\centering
\includegraphics[width=1.0\columnwidth]{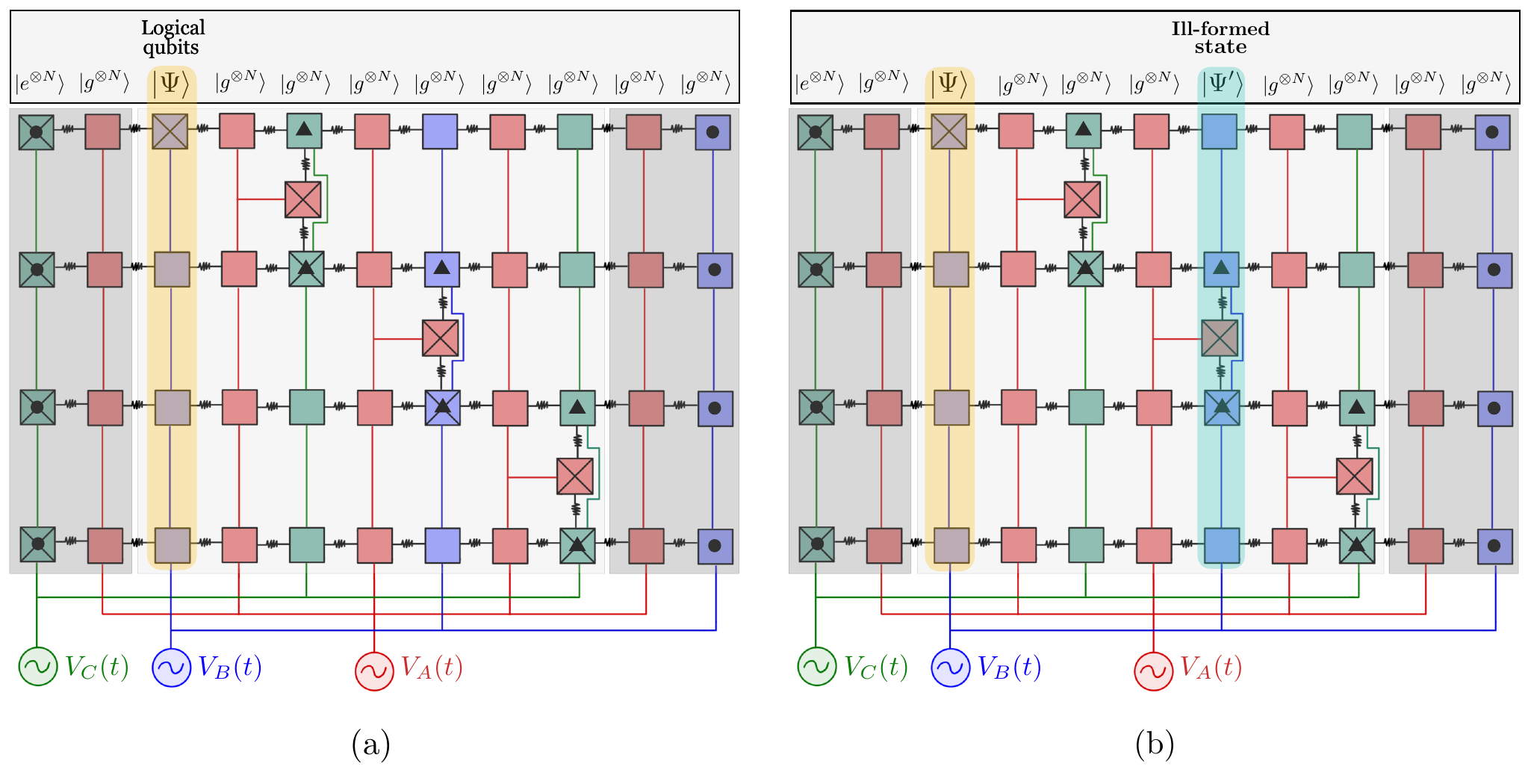}
\caption{Panel (a): example of a $B$-type well-formed state~(\ref{coding}) which we use to encode the logical information in the device
that allows the computation with $N=4$ logical qubits. 
The yellow area identifies the Information Carrying Column (ICC): the columns on the left of the ICC are in a paramagnetic phase, while
the columns on the right are in a ferromagnetic phase. The crossed red elements that act as couplers for the various rows of the ladder are
also in ferromagnetic ($|gg..g\rangle$) phase. Panel (b): example of a well-behaved, ill-formed state $|\tilde{\Psi};k\rangle$ obtained by deteriorating the state of panel (a) by replacing one 
 of the $B$ columns on the right-hand-side with a non-ferromagnetic term: under 
$\hat{Z}_{A^{\rm r}}^{\rm tot}$ this state transforms  as in Eq.~(\ref{ztotapp112233}). 
\label{fig:well}}
\end{figure}

As discussed in the main text, our model encodes  the information in one of the columns of the device.
Specifically at each step of the computation, we assume the ladder to be in a state of the form 
\begin{equation}\label{coding} |\Psi; {k}\rangle : =\Big( \cdots  |g^{\otimes N} \rangle_{k-3} \otimes |e^{\otimes N}\rangle_{k-2} \otimes |g^{\otimes N}\rangle_{k-1} \Big)  \otimes
|\Psi\rangle_k   \otimes 
\Big(   |g^{\otimes N} \rangle_{k+1}  \otimes  |g^{\otimes N} \rangle_{k+2} \otimes 
|g^{\otimes N} \rangle_{k+3}    \cdots \Big)\otimes |g^{\otimes (N-1)}\rangle_{A^\times}\;.
  \end{equation} 
In this expression  the ket $|\cdots \rangle_{k'}$ describes  the state of the $k'$-th column of the device: 
in particular $|g^{\otimes N}\rangle_{k'}$ ($|e^{\otimes N}\rangle_{k'}$) represents  the case
where all the qubits of the $k'$-th column are in the $|g\rangle$ ($|e\rangle$) configuration.
$|\Psi\rangle_k$ is the logical (possibly entangled) state of the ICC, and $|g^{\otimes (N-1)}\rangle_{A^\times}$ describes the state of the $N-1$ crossed $A$-type qubits which act as connectors between adjacent 
rows (notice that in writing  Eq.~(7) of the main text the presence of the last contribution was dropped for ease of  notation).
It is also worth stressing that, in our model, these are the only crossed elements of $A$-type; in particular, all the $A$ columns of the ladder are composed of only regular (non-crossed) elements.
\\ \\
States as defined in Eq.~(\ref{coding}) are said to be {\it well-formed} (see panel (a) of Fig.~\ref{fig:well} for an example). 
These states exhibit no quantum correlations among the different qubits of the device, 
except for those located in the ICC, which are the only ones that can share entanglement.
An alternative, more compact way to express them is:
\begin{equation}\label{codingpara} |\Psi; {k}\rangle =|{\rm para} \rangle \otimes
|\Psi\rangle_k   \otimes |{\rm ferro}\rangle\;,
\end{equation} 
which emphasises that the ICC is an interface between  a paramagnetic region on  the left, where, 
starting from the rightmost element which is always in $|g^{\otimes N}\rangle$,
all the columns are  prepared in an alternating sequence of  $|g^{\otimes N}\rangle$  and $|e^{\otimes N}\rangle$ configurations, and a ferromagnetic region  on the right where all the columns are instead  in the $|g^{\otimes N}\rangle$
configuration. In this compact notation, the ferromagnetic region always includes also the crossed qubits of type $A$.
The procedure  for initializing the ladder in this special type of states will be discussed in Sec.~\ref{sec:iniz}. Here, instead, we
discuss how the well-formed state of the model evolves under the action of the system Hamiltonian.
\\

{\bf Remark:} To simplify the presentation in the following Section, we shall introduce the special notation
``$\stackrel{\star}{=}$" to indicate that two vector states of the system are equal up to an (irrelevant, not state dependent) global phase. Specifically:
\begin{eqnarray} 
|a\rangle \stackrel{\star}{=} |b\rangle \qquad \Longrightarrow \quad \exists\;  e^{i\varphi}\;  \mbox{s.t.} \quad |a\rangle = e^{i\varphi}  |b\rangle\;. 
\end{eqnarray} 

\subsection{Preliminary observations} \label{sec:pre} 
To begin with, let us note a few useful properties.
\begin{itemize}
\item[i)] In our analysis we will only use states $|\Psi;k\rangle$ with values of $k$ that are larger than or equal to $3$ 
(i.e., the first column of the processing area of the device).
In other words, the logical information will always be located either in the processing area or in the read-out area, never in the initialization area.
 Additionally, the $|\Psi;k\rangle$'s can be categorized into three
groups: 
\begin{itemize}
\item {\bf $A$-type well-formed states:} these are states $|\Psi;k\rangle$  where the ICC index $k$ identifies a column of the ladder formed by $A$-type qubits; 
\item {\bf $B$-type well-formed states:} these are states  $|\Psi;k\rangle$  where the ICC index $k$ identifies a column of the ladder formed by $B$-type qubits; 
\item {\bf $C$-type well-formed states:} these are states  $|\Psi;k\rangle$  where the ICC index $k$ identifies a column of the ladder formed by $C$-type qubits.
\end{itemize} 
\item[ii)] The well-formed states $|\Psi;k\rangle$ are invariant under the action of the global unitary
$e^{-\frac{i}{\hbar}  \hat{H}_{\rm free} \tau_{\rm tot}}$ which enters in the decomposition~(\ref{stringSUP}). Specifically,
\begin{eqnarray}\label{prop1} 
e^{-\frac{i}{\hbar}  \hat{H}_{\rm free} \tau_{\rm tot}}|\Psi;k\rangle=e^{-2i \zeta \tau_{\rm tot}  \sum_{\langle i,j \rangle} \vert e_i e_{j} \rangle \langle e_i e_{j} \vert  }|\Psi;k\rangle = |\Psi;k\rangle\;. 
\end{eqnarray} 
This is due to the special arrangement of the interactions in the ladder, where in the state $|\Psi;k\rangle$, each ZZ-interacting
pair $\langle i,j\rangle$ has at least one element that is in the ground state, so  $\langle e_i e_j| \Psi;k\rangle=0$. 
Thanks to Eq.~(\ref{prop1}), and recalling that $e^{-\frac{i}{\hbar}  \hat{H}_{\rm free} \tau_{\rm tot}}$ commutes with all the 
$\hat{U}_{\chi}^{(\ell)}$ terms, we can write 
 \begin{eqnarray} \label{SstringSUPappli} \hat{U}_{\rm tot} \Big|_{\eta_{\rm BR}\gg 1} |\Psi;k\rangle &\simeq & 
 \hat{U}_{\chi}^{(\ell)} \cdots \hat{U}_{\chi}^{(2)} \; \hat{U}_{\chi}^{(1)}|\Psi;k\rangle
 \;, \end{eqnarray} 
which justifies dropping $e^{-\frac{i}{\hbar}  \hat{H}_{\rm free} \tau_{\rm tot}}$ in Eq.~(\ref{firstfirst}) and in Eq.~(5) of the main text. 
 \item[iii)]  Let $|\Psi;k\rangle$ be a $B$- or $C$-type well-formed state.
The application of the gate $\hat{Z}_{A^{\rm r}}^{\rm tot}$ of Eq.~(\ref{defZTOT}) induces the following mapping:
\begin{eqnarray} 
\hat{Z}_{A^{\rm r}}^{\rm tot} |\Psi;k\rangle \stackrel{\star}{=}  
\left|[\hat{\sigma}^{(z)}]^{\otimes N} \Psi;k\right\rangle
:= 
|{\rm para} \rangle \otimes [\hat{\sigma}^{(z)}]^{\otimes N} |\Psi\rangle_k  \otimes |{\rm ferro}\rangle\;,\label{ztotapp} 
\end{eqnarray} 
where $\left|[\hat{\sigma}^{(z)}]^{\otimes N} \Psi;k\right\rangle$ represents a well-formed state of the system with the  ICC still located in the $k$-th
column of the device, but with the internal state of the logical qubits evolved under the action of a contemporary application of  $\hat{\sigma}^{(z)}$ gates. To see why Eq.~(\ref{ztotapp}) holds, recall that $\hat{Z}_{A^{\rm r}}^{\rm tot}$ acts selectively on the regular $A$-type qubits of the model, applying to each of them a   control-$2\pi$ rotation, i.e. a control-phase shift  of $-1$. Observes that, since the ICC column is of $B$- or $C$-type, then the only $A$-type columns that can acquire such extra phase, are those in the ferromagnetic area. Indeed all the $A$-type columns on the left-hand-side of the ICC have at least either an interacting neighbouring $B$ or $C$ element that is in the excited state. Apart form the first one, the   $A$-type columns in the ferromagnetic region  are coupled with neighbouring $B$- and $C$-type qubits which are in a definite ground state $|g\rangle$ -- see panel (a) of Fig.~\ref{fig:well}. Acting on them, $\hat{Z}_{A^{\rm r}}^{\rm tot}$ will assign to the state an irrelevant global phase that is a product of $(-1)$'s. The exact number  $M_k$ of such terms depends on the actual location of ICC column. For instance in the case shown in the panel (a) of Fig.~\ref{fig:well}, the global phase induced by $\hat{Z}_{A^{\rm r}}^{\rm tot}$ is $(-1)^{\cdots}=\cdots$. 
In the end the only non trivial effect that $\hat{Z}_{A^{\rm r}}^{\rm tot}$ can introduce in the model occurs when it acts on the first $A$-type column of the ferromagnetic area.
For future reference we shall identify such column with the symbol ${\cal A}^{\rm r}_*$. Due to the presence of the ICC, the activation of the $(-1)$ phase on the $j$-th row  element of this column depends on the state of the  $j$-th logical qubit (observe that the only other $B$- or $C$-type column that is ZZ coupled with ${\cal A}^{\rm r}_*$ has all the qubits in the $|g\rangle$ state).
To formalize this, define $\hat{Z}_{{\cal A}^{\rm r}_*}^{\rm tot}$ as the restriction of $\hat{Z}_{A^{\rm r}}^{\rm tot}$ on the column ${\cal A}^{\rm r}_*$, i.e.
\begin{eqnarray}
\hat{Z}_{{\cal A}^{\rm r}_*}^{\rm tot}:=  \prod_{i\in {\cal A}^{\rm r}_*}   \left[  \hat{Q}_{\langle i \rangle} -   \hat{P}_{\langle i \rangle} \right] \;.
\end{eqnarray}  
From the previous analysis, we can write: 
 \begin{eqnarray} 
 \hat{Z}_{A^{\rm r}}^{\rm tot} |\Psi;k\rangle = (-1)^{M_k} \hat{Z}_{{\cal A}^{\rm r}_*}^{\rm tot} |\Psi;k\rangle = 
 (-1)^{M_k} \prod_{i\in {\cal A}^{\rm r}_*}    \left[  \hat{Q}_{\langle i \rangle} -   \hat{P}_{\langle i \rangle} \right] |\Psi;k\rangle \;.
 \label{ztotapp11} 
 \end{eqnarray} 
Notice  that $\hat{Q}_{\langle i \rangle} -   \hat{P}_{\langle i \rangle}$ is a control-phase gate that act on the $i$-th qubit of ${\cal A}^{\rm r}_*$, 
which is activated if and only if its two  ZZ coupled neighbouring qubits are both in the ground state. 
To make use of this fact expand the logical state $|\Psi\rangle$ of the ICC as
 \begin{eqnarray} \label{representation} 
 |\Psi\rangle = |g_i \Psi_g\rangle + |e_i \Psi_e\rangle  \qquad \Longrightarrow\qquad 
 |\Psi;k \rangle = |g_i \Psi_g;k \rangle + |e_i \Psi_e; k \rangle\;, 
 \end{eqnarray} 
where $|g_i\rangle$ and $|e_i\rangle$ are the ground and excited state of the element pertaining to the ICC column that is on the same row of $i$-th qubit of ${\cal A}^{\rm r}_*$, $|\Psi_g\rangle$ and $|\Psi_e\rangle$ are (non necessarily normalized) states of the remaining $N-1$ elements of the ICC column. Observe then that
\begin{eqnarray} 
\left\{ \begin{array}{l} 
\hat{Q}_{\langle i \rangle}|\Psi;k \rangle = |e_i \Psi_e; k \rangle\;, \\
\hat{P}_{\langle i \rangle}|\Psi;k \rangle = |g_i \Psi_g; k \rangle\;, 
\end{array} \right. \quad \Longrightarrow \quad (\hat{Q}_{\langle i \rangle} -   \hat{P}_{\langle i \rangle}) |\Psi;k \rangle= 
- |g_i \Psi_g;k \rangle + |e_i \Psi_e; k \rangle  = \hat{\sigma}^{(z)}_i |\Psi;k \rangle\;, \label{impo111} 
\end{eqnarray} 
 which when replaced in the previous Eq.~(\ref{ztotapp11}) gives
  \begin{eqnarray} 
 \hat{Z}_{A^{\rm r}}^{\rm tot} |\Psi;k\rangle =
 (-1)^{M_k}  \prod_{i\in {\cal A}^{\rm r}_*} \hat{\sigma}^{(z)}_i |\Psi;k \rangle = (-1)^{M_k} \left|[\hat{\sigma}^{(z)}]^{\otimes N} \Psi;k\right\rangle\;, 
 \label{ztotapp1111}
 \end{eqnarray} 
proving Eq.~(\ref{ztotapp}).
\item[iv)] The property~(\ref{ztotapp}) also holds for certain ill-formed states of the ladder. In particular  
this occurs if one or more of the $B$-type ($C$-type) columns in the ferromagnetic phase of a $B$-type ($C$-type) well-formed state $|\Psi;k\rangle$  are replaced by arbitrary qubit states (see panel (b) of Fig.~\ref{fig:well} for an example of a configuration that meets these requirements). 
 An ill-formed state of this type will be said to be {\it well-behaved}.
Under these conditions, each of the two $A$ columns near a deteriorated $B$ column 
of well-behaved, ill-formed $B$-type state 
has an extra $C$ column that is still initialized in the proper ferromagnetic  configuration $|g^{\otimes N}\rangle$. 
Thus, when $\hat{Z}_{A^{\rm r}}^{\rm tot}$ is applied to the system, a deteriorated
$B$ column will act as the unique controller for its two neighbouring $A$ elements, assigning the  same conditional $(-1)$ phase  to both,  ultimately not modifying the state.
A similar effect will occur in case of well-behaved, ill-formed $C$-type state: here 
each of the two $A$ columns near a deteriorated $C$ column will have an extra $B$ column that is still initialized in the proper ferromagnetic configuration $|g^{\otimes N}\rangle$.

To see this explicitly let us consider in details the case of a well-behaved, ill-formed $B$-type state. $|\tilde{\Psi};k\rangle$ (the analysis of the well-behaved, ill-formed $C$-type states will follows along the same path).
Define $\{ \bar{\cal B}_1, \bar{\cal B}_2, \cdots, \bar{\cal B}_S\}$ the set of its deformed columns. Let  $\bar{\cal A}_{\ell}^{({\rm left})}$ and $\bar{\cal A}_{\ell}^{({\rm right})}$ the $A$ columns on the left-hand side and right-hand side, respectively, of the $\ell$-th element of this set. As in the case of Eq.~(\ref{ztotapp1111}) we can write 
 \begin{eqnarray} 
 \hat{Z}_{A^{\rm r}}^{\rm tot}| \tilde{\Psi};k\rangle = (-1)^{M'_k} \hat{Z}_{{\cal A}^{\rm r}_*}^{\rm tot} \prod_{\ell=1}^S
 \hat{Z}_{\bar{\cal A}_{\ell}^{({\rm left})}}^{\rm tot} \hat{Z}_{\bar{\cal A}_{\ell}^{({\rm right})}}^{\rm tot}  |\tilde{\Psi};k\rangle\;, 
 \label{ztotapp1122} 
 \end{eqnarray} 
where the global phase $(-1)^{M'_k}$ depends on the number ${M'_k}$ of ferromagnetic $B$ columns that are not deteriorated. Here 
$\hat{Z}_{{\cal A}}^{\rm tot}:=  \prod_{i\in {\cal A}}   \left[  \hat{Q}_{\langle i \rangle} -   \hat{P}_{\langle i \rangle} \right]$ indicates the part of $\hat{Z}_{A^{\rm r}}^{\rm tot}$ that involves elements of the ${\cal A}$ column. Due to the commutativity of the operators $\left[  \hat{\openone}_i \otimes \hat{Q}_{\langle i \rangle}+
\hat{\mathbb{R}}_i(\theta,\bm{n}) \otimes \hat{P}_{\langle i \rangle}  \right]$ associated with different qubits, 
$\hat{Z}_{\cal A}^{\rm tot}$ of different columns commute. 
Following the same derivation used  in Eq.~(\ref{ztotapp1111}), we then can show that 
\begin{eqnarray} 
\hat{Z}_{\bar{\cal A}_{\ell}^{({\rm left})}}^{\rm tot} \hat{Z}_{\bar{\cal A}_{\ell}^{({\rm right})}}^{\rm tot} 
 |\tilde{\Psi};k\rangle = \hat{Z}_{\bar{\cal A}_{\ell}^{({\rm left})}}^{\rm tot} \left( \prod_{i\in \bar{\cal B}_\ell} \hat{\sigma}^{(z)}_i |\tilde{\Psi};k\rangle\right)
 = \prod_{i'\in \bar{\cal B}_\ell} \hat{\sigma}_{i'}^{(z)}  \prod_{i\in \bar{\cal B}_\ell} \hat{\sigma}^{(z)}_i |\tilde{\Psi};k\rangle =  |\tilde{\Psi};k\rangle\;,
\end{eqnarray} 
so that 
\begin{eqnarray} 
 \hat{Z}_{A^{\rm r}}^{\rm tot}| \tilde{\Psi};k\rangle = (-1)^{M_k'} \hat{Z}_{{\cal A}^{\rm r}_*}^{\rm tot}  |\tilde{\Psi};k\rangle=
  (-1)^{M_k'} \left|[\hat{\sigma}^{(z)}]^{\otimes N} \tilde{\Psi};k\right\rangle\;, 
 \label{ztotapp112233} 
 \end{eqnarray} 
 which proves the thesis. 
 \end{itemize} 
 
\subsection{Motion of the interface}\label{sec:motioninter} 
As we have mentioned in the main text, by applying a specific sequence of pulses~(\ref{SstringSUP}) we can generate
a unitary transformation $\hat{U}_{\rm shift}$ that  rigidly moves the interface from left to right, i.e. 
\begin{eqnarray} \label{equshift} 
\hat{U}_{\rm shift} |\Psi; {k}\rangle \stackrel{\star}{=}   |\Psi; {k + 1}\rangle\;,
\end{eqnarray} 
for all $3\leq k \leq  2N$ (recall that the total numbers of column in the model is $2N+3$).  
 In particular given 
\begin{eqnarray} \label{dfdsfadf} 
\hat{\Pi}_{\xi}:= \hat{W}_{\xi}( \pi ; \bm{x})=
 \prod_{i\in {\xi}}   \left[  \hat{\openone}_i \otimes \hat{Q}_{\langle i \rangle}- i \hat{\sigma}^{(x)}_i\otimes    \hat{P}_{\langle i \rangle} \right]\;, 
\end{eqnarray} 
this can be done e.g. by taking
\begin{eqnarray}\label{ushift} 
\hat{U}_{\rm shift} = \left\{ \begin{array}{lll} \hat{\Pi}_{A^{\rm r}}
\hat{\Pi}_{B}\hat{\Pi}_{C}\hat{\Pi}_{A^{\rm r}} && \mbox{if  $|\Psi; {k}\rangle$  is either a $B$ or $C$-type well-formed state;}\\ \\
\hat{\Pi}_{B}\hat{\Pi}_{C}\hat{\Pi}_{A^{\rm r}} \hat{\Pi}_{B}\hat{\Pi}_{C}&&  \mbox{if  $|\Psi; {k}\rangle$  is a $A$-type well-formed state,} \end{array} \right. 
\end{eqnarray}
which we write omitting the term $e^{-\frac{i}{\hbar}  \hat{H}_{\rm free} \tau_{\rm tot}}$ due to property ii) of Sec.~\ref{sec:pre}.
To verify that our choice of $\hat{U}_{\rm shift}$ is correct, consider  the case where $|\Psi; {k}\rangle$  is a $B$-type well-formed state (the analysis for the other cases  is similar and will not be reported here). 
To study the evolution of this state under the sequence $\hat{\Pi}_{A^{\rm r}}
\hat{\Pi}_{B}\hat{\Pi}_{C}
\hat{\Pi}_{A^{\rm r}}$ it is worth splitting 
the vector $|\Psi; {k}\rangle$ in the following four sectors:
\begin{eqnarray}\label{codingSPLIT} |\Psi; {k}\rangle&: =&\Big( \underbrace{\cdots |g_A^{\otimes N} \rangle_{k-5} \otimes
|e_B^{\otimes N}\rangle_{k-4} \otimes  |g_A^{\otimes N} \rangle_{k-3} \otimes |e_C^{\otimes N}\rangle_{k-2} \otimes |g_A^{\otimes N}\rangle_{k-1} }_{\mbox{sector I}} \Big)  \otimes \Big(\underbrace{|\Psi_B\rangle_k  \otimes
  |g_A^{\otimes N} \rangle_{k+1}}_{\mbox{sector II}}\Big) \nonumber \\
  && \otimes \Big(  \underbrace{|g_C^{\otimes N} \rangle_{k+2} \otimes 
|g_A^{\otimes N} \rangle_{k+3} \otimes |g_B^{\otimes N} \rangle_{k+4} \otimes 
|g_A^{\otimes N} \rangle_{k+5}    \cdots }_{\mbox{sector III}} \Big)\otimes |\underbrace{g^{\otimes (N-1)}}_{\mbox{sector IV}} \rangle_{A^\times}
  \end{eqnarray} 
  where we add the subscripts $A$, $B$, and $C$ to the various terms to specify the type of the column. 
 \begin{itemize} 
\item{\bf Sector IV}:  First of all, it is clear that  since the crossed $A$-type qubits are not affected by the selected controls, 
 the term in sector IV will be left unchanged by the evolution, i.e. 
\begin{eqnarray} |g^{\otimes (N-1)}\rangle_{A^\times} \stackrel{\hat{\Pi}_{A^{\rm r}}
\hat{\Pi}_{B}\hat{\Pi}_{C}
\hat{\Pi}_{A^{\rm r}}}{\longrightarrow} |g^{\otimes (N-1)}\rangle_{A^\times} \;. \label{sectorIV} 
\end{eqnarray} 
Most importantly, it is worth stressing that the crossed $A$ elements will remain in the $|g\rangle$ state at {\it all times} during the entire
pulse sequence. This fact is extremely important as it  means that 
these qubits play no role what-so-ever in the evolution of the other qubits of the device. In particular the possibility that a given control-pulse
will be activated on one of the qubits that are ZZ coupled with a crossed $A$-type element, does not depend on the latter since it will be always in the 
``go" state. 

\item{\bf Sector I}:  According to Eq.~(\ref{codingSPLIT}) the qubits of this sector are initiliazed in the paramagnetic phase. 
As evident from the equation in this case the $A$ columns have at least a nearby  $B$ or $C$ column  in the $|e^{\otimes N}\rangle$ which prevents the first $\hat{\Pi}_{A^{\rm r}}$  of the sequence to operate on them.  Viceversa, when we apply $\hat{\Pi}_{B}$, since all the $B$ columns have neighbouring $A$-type qubits which are all in $|g\rangle$, they will experience 
$-i \hat{\sigma}_i^{(x)}$ rotations that will map them into $(-i)^{N}|g^{\otimes N}\rangle$. The final $\hat{\Pi}_{A^{\rm r}}$ operation
will hence not be blocked, modifying the states of the $A$-type qubits in the paramagnetic area via $-i \hat{\sigma}_i^{(x)}$ pulses.
The same happens to the $C$ columns when we operate with $\hat{\Pi}_{C}$ (recall that due to the commutative property Eq.~(\ref{COMMBC})
the order in which we apply $\hat{\Pi}_{B}$ and $\hat{\Pi}_{C}$ does not matter). The global trajectory induced by the unitary $\hat{U}_{\rm shift}$ on the component (\ref{primad}) of $|\Psi;k\rangle$ can hence be expressed as
\begin{eqnarray}
 \cdots 
|e_B^{\otimes N}\rangle_{k-4}   \otimes |g_A^{\otimes N} \rangle_{k-3} \otimes |e_C^{\otimes N}\rangle_{k-2} \otimes |g_A^{\otimes N}\rangle_{k-1} &\stackrel{\hat{\Pi}_{A^{\rm r}}}{\longrightarrow}& \cdots 
|e_B^{\otimes N}\rangle_{k-4}   \otimes |g_A^{\otimes N} \rangle_{k-3} \otimes |e_C^{\otimes N}\rangle_{k-2} \otimes |g_A^{\otimes N}\rangle_{k-1}\nonumber \\ 
 &\stackrel{\hat{\Pi}_{B}}{\longrightarrow}& \cdots  
|g_B^{\otimes N}\rangle_{k-4}\otimes |g_A^{\otimes N} \rangle_{k-3} \otimes |e_C^{\otimes N}\rangle_{k-2} \otimes |g_A^{\otimes N}\rangle_{k-1} \nonumber 
\\
&\stackrel{\hat{\Pi}_{C}}{\longrightarrow}& \cdots  
|g_B^{\otimes N}\rangle_{k-4}\otimes |g_A^{\otimes N} \rangle_{k-3} \otimes |g_C^{\otimes N}\rangle_{k-2} \otimes |g_A^{\otimes N}\rangle_{k-1} \nonumber \\
&\stackrel{\hat{\Pi}_{A^{\rm r}}}{\longrightarrow}& \cdots
|g_B^{\otimes N}\rangle_{k-4}\otimes |e_A^{\otimes N} \rangle_{k-3} \otimes |g_C^{\otimes N}\rangle_{k-2} \otimes |e_A^{\otimes N}\rangle_{k-1}\;,
\nonumber \\ \label{primad} 
\end{eqnarray} 
up to an irrelevant global phase shift given by $(-i)^{M_k}$  with $M_k$ counting the total number of the applied $-i \hat{\sigma}_i^{(x)}$ rotations 
(a number that depends only on $k$). 
\item{\bf Sector III}: 
In this case it is clear that the first $\hat{\Pi}_{A^{\rm r}}$ can induce $-i \hat{\sigma}_i^{(x)}$ rotation on all
the $A$ columns, bringing them into $|e^{\otimes N}\rangle$ states which, in turn, will  prevent the action of 
$\hat{\Pi}_{B}$ and $\hat{\Pi}_{C}$. The final operator $\hat{\Pi}_{A^{\rm r}}$ will still be able to act returning all the $A$ column to their original configuration, i.e. 
\begin{eqnarray}
|g_C^{\otimes N} \rangle_{k+2} \otimes 
|g_A^{\otimes N} \rangle_{k+3}  \otimes |g_B^{\otimes N} \rangle_{k+4} \otimes |g_A^{\otimes N} \rangle_{k+5}
  \cdots  &\stackrel{\hat{\Pi}_{A^{\rm r}}}{\longrightarrow}& 
|g_C^{\otimes N} \rangle_{k+2} \otimes 
|e_A^{\otimes N} \rangle_{k+3}  \otimes |g_B^{\otimes N} \rangle_{k+4} \otimes |e_A^{\otimes N} \rangle_{k+5}
  \cdots  \nonumber \\ 
 &\stackrel{\hat{\Pi}_{B}}{\longrightarrow}& 
|g_C^{\otimes N} \rangle_{k+2} \otimes 
|e_A^{\otimes N} \rangle_{k+3}  \otimes |g_B^{\otimes N} \rangle_{k+4} \otimes |e_A^{\otimes N} \rangle_{k+5}
  \cdots
\nonumber 
\\
&\stackrel{\hat{\Pi}_{C}}{\longrightarrow}& 
|g_C^{\otimes N} \rangle_{k+2} \otimes 
|e_A^{\otimes N} \rangle_{k+3}  \otimes |g_B^{\otimes N} \rangle_{k+4} \otimes |e_A^{\otimes N} \rangle_{k+5}
  \cdots
\nonumber 
\\
&\stackrel{\hat{\Pi}_{A^{\rm r}}}{\longrightarrow}& 
|g_C^{\otimes N} \rangle_{k+2} \otimes 
|g_A^{\otimes N} \rangle_{k+3}  \otimes |g_B^{\otimes N} \rangle_{k+4} \otimes |g_A^{\otimes N} \rangle_{k+5}
  \cdots \;. \nonumber \\\label{ultimad} 
\end{eqnarray} 
\item{\bf Sector II}: 
The study of this sector  is slightly more complex since in this case the various pulses induce correlations among involved columns (i.e. the ICC column and its first neighbour on the right). To begin with it is clear that 
$\hat{\Pi}_{C}$ will not have effects on the sector as it does not include $C$ qubits.
A useful observation is also that 
when we apply the  two $\hat{\Pi}_{A^{\rm r}}$ transformations of the sequence, the action of these operators on the $(k+1)$-th column (which is of $A$-type), only
depends on the internal state of the ICC column, due to the fact that  the $C$ column at position $k+2$ is in the state $|g^{\otimes N}\rangle$ -- see
Eq.~(\ref{ultimad}). 
Similarly when we apply the $\hat{\Pi}_{B}$, its action on the ICC column only depends on the $(k+1)$-th column, due to the fact that 
the $A$ column at position $k-1$ is in $|g^{\otimes N}\rangle$ as shown in  Eq.~(\ref{primad}) (recall that the crossed $A$-type qubits are always in the $|g\rangle$ during the entire process). We can hence conclude that, despite our control pulses affect more than two qubits at the time, thanks to the selected encoding, the net effect of the sequence  $\hat{\Pi}_{A^{\rm r}}
\hat{\Pi}_{B}\hat{\Pi}_{A^{\rm r}}$ is to effectively  induce a selective coupling between column $k$ with column $k+1$. 
To study explicitly what type of evolution this induces in our model, it is useful to  consider first the scenario where the ladder is formed by a single row. 
In this case the state of sector II can be explicitly expressed as 
\begin{eqnarray}|\Psi_B\rangle_k\otimes 
  |g_A \rangle_{k+1}  =  \alpha |g_B \rangle_k \otimes  |g_A \rangle_{k+1}    + \beta |e_B\rangle_k \otimes  |g_A \rangle_{k+1}   \;, 
 \end{eqnarray} 
 where we expanded $|\Psi_B\rangle_k$ in the computational basis. 
 We can now easily track the evolution of this configuration recalling that the various control-operations are activated either by the internal state of
 $k$-th column (in the case of $\hat{\Pi}_{A^{\rm r}}$), or by the internal state of the $(k+1)$-th column 
 (in the case of $\hat{\Pi}_{B}$). Accordingly we get:
 \begin{eqnarray}\label{ultimadd} 
|\Psi_B\rangle_k\otimes 
  |g_A \rangle_{k+1}  &\stackrel{\hat{\Pi}_{A}}{\longrightarrow}& 
-i  \alpha |g_B \rangle_k \otimes  |e_A \rangle_{k+1}    + \beta |e_B\rangle_k \otimes  |g_A \rangle_{k+1}
 \nonumber \\ 
 &\stackrel{\hat{\Pi}_{B}}{\longrightarrow}& 
-i  \alpha |g_B \rangle_k \otimes  |e_A \rangle_{k+1}    - i \beta |g_B\rangle_k \otimes  |g_A \rangle_{k+1}=
-i |g_B \rangle_k \otimes  (\alpha   |e_A \rangle_{k+1}    + \beta  |g_A \rangle_{k+1}) \nonumber 
\\
&\stackrel{\hat{\Pi}_{A}}{\longrightarrow}& 
(-i)^2  |g_B \rangle_k \otimes  (\alpha   |g_A \rangle_{k+1}    + \beta  |e_A \rangle_{k+1}) = - |g_B \rangle_k \otimes 
|\Psi_A\rangle_{k+1}\;,
\end{eqnarray} 
which shows that at the end of the sequence the states of the $k$ and $k+1$ have swapped (up to an irrelevant global phase). 
To generalize this result to the case of $N$ rows, simply recall that in our model, apart from the presence of the crossed $A$-type qubits which in the present case are always in the $|g\rangle$ state, there are not direct interactions among the various elements of a column. This in particular implies that the dynamics induced by $\hat{\Pi}_{A^{\rm r}}
\hat{\Pi}_{B}\hat{\Pi}_{A^{\rm r}}$ can be addressed treating the various rows independently. Accordingly we can conclude that (up to a global phase) one has, 
\begin{eqnarray}\label{sectorII} 
|\Psi_B\rangle_k  \otimes
  |g_A^{\otimes N} \rangle_{k+1}  \stackrel{\hat{\Pi}_{A^{\rm r}}
\hat{\Pi}_{B}\hat{\Pi}_{C}
\hat{\Pi}_{A^{\rm r}}}{\longrightarrow}
  |g_B^{\otimes N} \rangle_{k} \otimes  |\Psi_A\rangle_{k+1}   \;.
\end{eqnarray} 
\end{itemize} 
The proof of Eq.~(\ref{equshift}) finally follows by putting together  all the identities obtained for the different sectors (again the result is obtained up to an irrelevant global phase which does not depend on the input state of the ICC). \\
We finally remark that an alternative (yet fully equivalent) implementation of the transformation~(\ref{equshift}) could have been realized by replacing all the $\pi$ angles appearing  in Eq.~(\ref{dfdsfadf}) with $-\pi$'s (with this choice the state $|\Psi; k\rangle$ will still mapped
into $|\Psi; k+1\rangle$ up to a global phase). This innocent looking observation is useful to clarify why the same operators given in 
Eq.~(\ref{ushift}) will also induce the reverse of the mapping~(\ref{equshift}), i.e. 
\begin{eqnarray} \label{equshiftrev} 
\hat{U}_{\rm shift} |\Psi; {k+1}\rangle \stackrel{\star}{=}  |\Psi; {k}\rangle\;.
\end{eqnarray} 
Indeed  changing the sign of the  $\pi$'s in Eq.~(\ref{dfdsfadf}) means taking the inverse of $\hat{\Pi}_{A^{\rm r}}$, $\hat{\Pi}_{B}$ and $\hat{\Pi}_{C}$. Therefore for the case where $|\Psi;k\rangle$ is a $B$-type well-formed state we can write 
 \begin{eqnarray} 
\hat{\Pi}_{A^{\rm r}}
\hat{\Pi}_{B}\hat{\Pi}_{C}
\hat{\Pi}_{A^{\rm r}} |\Psi; {k}\rangle &\stackrel{\star}{=}&   \hat{\Pi}^\dag_{A^{\rm r}}
\hat{\Pi}^\dag_{C}\hat{\Pi}^\dag_{B}
\hat{\Pi}^\dag_{A^{\rm r}} |\Psi; {k}\rangle \stackrel{\star}{=}   |\Psi; {k+1}\rangle \nonumber \\
&& \Longrightarrow \quad 
\left( \hat{\Pi}_{A^{\rm r}}
\hat{\Pi}_{B} \hat{\Pi}_{C}
\hat{\Pi}_{A^{\rm r}}\right)^2 |\Psi; {k}\rangle \stackrel{\star}{=}     |\Psi; {k}\rangle \stackrel{\star}{=}  \hat{\Pi}_{A^{\rm r}}
\hat{\Pi}_{B}\hat{\Pi}_{C}
\hat{\Pi}_{A^{\rm r}} |\Psi; {k+1}\rangle \;, \label{ushiftres}
\end{eqnarray} 
which proves the thesis (for $A$- and $C$-type well-formed states we can proceed similarly).

\subsubsection{Explicit pulse sequence} 
Notice that  $\hat{\Pi}_{B}$ induces $\pi$-pulses on all the $B$-type qubits (crossed and regular). A specific instance of the control parameters that allows for such operation is obtained by considering the following four step sequence $\hat{U}^{(4)}_B\hat{U}^{(3)}_{B} \hat{U}^{(2)}_{B}\hat{U}^{(1)}_{B}$ where 
\begin{itemize}
    \item $\hat{U}^{(1)}_{B}$ is induced by setting $\phi_B^{(1)} = 0$ and letting the system evolve for a time $\tau_1= \frac{3\pi}{4}\Omega_B^{-1}$;
    \item  $\hat{U}^{(2)}_{B}$ is induced by setting $\phi^{(2)}_B = \pi/2$ and letting the system evolve
    for a time $\tau_2= \pi\Omega_B^{-1}$; 
    \item  $\hat{U}^{(3)}_{B}$ is induced by setting $\phi^{(3)}_B = \pi$ and letting the system evolve
    for a time $\tau_3= \frac{\pi}{4}\Omega_B^{-1}$;
    \item $\hat{U}^{(4)}_{B}$ is induced by setting  $\phi^{(4)}_B = -\pi/2$ and letting the system evolve
    for a time $\tau_4= \pi\Omega_B^{-1}$;
\end{itemize}
(recall that by convention $\Omega_B$ is the Rabi frequency of the regular $B$-type qubits).
A similar decomposition applies also to $\hat{\Pi}_{C}$. 
On the contrary $\hat{\Pi}_{A^{\rm r}}$ induces $\pi$-pulses only the regular elements of the $A$ qubits. A specific instance of the control parameters
that allows for such operation is obtained by considering the following four step sequence 
$\hat{U}^{(4)}_{A}\hat{U}^{(3)}_{A} \hat{U}^{(2)}_{A}\hat{U}^{(1)}_{A}$
where now
\begin{itemize}
    \item  $\hat{U}^{(1)}_{A}$ is induced by setting $\phi_A^{(1)} = 0$ and letting the system evolve 
    for a time $\tau_1= \frac{\pi}{2} \Omega_A^{-1}$;
    \item $\hat{U}^{(2)}_{A}$ is induced by setting $\phi_A^{(2)} = \pi/2$ and letting the system evolve 
    for a time $\tau_2 = \pi \Omega_A^{-1}$; 
    \item $\hat{U}^{(3)}_{A}$ is induced by setting $\phi_A^{(3)} = \pi$ and letting the system evolve 
    for a time $\tau_3 = \frac{\pi}{2} \Omega_A^{-1}$;
    \item $\hat{U}^{(4)}_{A}$ is induced by setting $\phi_A^{(4)} = -\pi/2$ and letting the system evolve 
    for a time $\tau_4= \pi \Omega_A^{-1}$;
\end{itemize}
(also $\Omega_A$ is the Rabi frequency of the regular $A$-type qubits).

\subsection{Single-qubit gate}\label{sec:single-qubit}
Referring to the previous subsection, we are able to move the interface in whatever desired position of the ladder, specifically where a $\chi$-type crossed-qubit is positioned with $\chi$ being either equal to $B$ or $C$. In order to perform a single-qubit gate, once the interface is located at the $\chi$-type crossed-qubit,  we need to send a specific global sequence of pulses involving $A$- and $\chi$-type qubits. 
Specifically the 
operations needed for implementing the single-qubit gates can be realized composing control pulses that induce evolutions 
of the form 
\begin{eqnarray} 
\hat{Z}_{A^{\rm r}}^{{\rm tot}} \hat{W}_{\chi^{\times}}(\theta/2;-\bm{n}_\perp) 
\hat{Z}_{A^{\rm r}}^{{\rm tot}} \hat{W}_{\chi^{\times}}(\theta/2;\bm{n}_\perp)
\end{eqnarray} 
with $\bm{n}_\perp$ orthogonal to $\bm{z}$ and 
$\hat{Z}_{A^{\rm r}}^{{\rm tot}}$  defined in Eq.~(\ref{defZTOT}). 
Indeed thanks to the fact that the crossed $B$-type qubits are located on columns which are
at least three columns apart from each other, when acting on $|\Psi; {k}\rangle$ the above transformation will effectively 
correspond to apply the single-qubit rotation  $\hat{\mathbb{R}}(\theta,\bm{n}_\perp)$ on the crossed element of the
ICC. 
The key observations here are the properties iii) and iv) of Sec.~\ref{sec:pre}. 
Notice in fact that since $|\Psi,k\rangle$ is a well-formed state of $\chi$-type, with $k$ being associated with 
a column $\chi$ that contains a crossed term, after the action of 
$\hat{W}_{\chi^{\times}}(\theta/2;\bm{n}_\perp)$ it will become a well-behaved,  ill-formed state,
 \begin{eqnarray}\label{prima1} 
\hat{W}_{\chi^{\times}}(\theta/2;\bm{n}_\perp) |\Psi;k\rangle = |\tilde{\Psi};k\rangle :=
\bigotimes_{{i\in \chi^{\times}/{\rm PARA}}} \hat{\mathbb{R}}_i(\theta/2,\bm{n}_\perp)  |\Psi;k\rangle
 \;, 
\end{eqnarray} 
where in the last identity we emphasize that  all the crossed $\chi$
qubits that are not in the paramagnetic area (including the one in the ICC) 
acquires the rotation $\hat{\mathbb{R}}_i(\theta/2,\bm{n}_\perp)$.
Therefore under 
$\hat{Z}_{A^{\rm r}}^{{\rm tot}}$ this state will evolve as in Eq.~(\ref{ztotapp112233}) 
acquiring  
$\hat{\sigma}^{(z)}$-gates  to each of the element of the ICC, i.e. explicitly 
\begin{eqnarray} \label{prima2} 
\hat{Z}_{A^{\rm r}}^{{\rm tot}}  \hat{W}_{\chi^{\times}}(\theta/2;\bm{n}_\perp) |\Psi;k\rangle = 
\hat{Z}_{A^{\rm r}}^{{\rm tot}}  |\tilde{\Psi};k\rangle= (-1)^{M_k'}  
\bigotimes_{\stackrel{i\in \chi^{\times}/{\rm PARA}}{j\in {\rm ICC}}} \hat{\sigma}_j^{(z)} \hat{\mathbb{R}}_i(\theta/2,\bm{n}_\perp)  |\Psi;k\rangle\;.
\end{eqnarray} 
The action of $\hat{W}_{\chi^{\times}}(\theta/2;-\bm{n}_\perp)$ will be similar to what seen in Eq.~(\ref{prima1}), i.e. 
\begin{eqnarray} 
\hat{W}_{\chi^{\times}}(\theta/2;-\bm{n}_\perp)  \hat{Z}_{A^{\rm r}}^{{\rm tot}}  \hat{W}_{\chi^{\times}}(\theta/2;\bm{n}_\perp) |\Psi;k\rangle = 
(-1)^{M_k'}  
\bigotimes_{\stackrel{i\in \chi^{\times}/{\rm PARA}}{j\in {\rm ICC}}
} \hat{\mathbb{R}}_{i}(\theta/2,-\bm{n}_\perp) \hat{\sigma}_j^{(z)} \hat{\mathbb{R}}_i(\theta/2,\bm{n}_\perp)  |\Psi;k\rangle\;.
\end{eqnarray} 
Since the latter is again well-behaved we can then replicate the argument of (\ref{prima2}) to write 
\begin{eqnarray} 
\hat{Z}_{A^{\rm r}}^{{\rm tot}}\hat{W}_{\chi^{\times}}(\theta/2;-\bm{n}_\perp)  \hat{Z}_{A^{\rm r}}^{{\rm tot}}  \hat{W}_{\chi^{\times}}(\theta/2;\bm{n}_\perp) |\Psi;k\rangle = 
(-1)^{2M_k'}  
\bigotimes_{\stackrel{i\in \chi^{\times}/{\rm PARA}}{j\in {\rm ICC}}
} \hat{\sigma}_j^{(z)} \hat{\mathbb{R}}_{i}(\theta/2,-\bm{n}_\perp) \hat{\sigma}_j^{(z)} \hat{\mathbb{R}}_i(\theta/2,\bm{n}_\perp)  |\Psi;k\rangle\;.
\end{eqnarray} 
Now the thesis follows by  observing that if $j$ corresponds to a non-crossed element of the ICC, it will experience the action of $\hat{\sigma}_j^{(z)} \hat{\sigma}_j^{(z)}=
\hat{\openone}_j$; similarly if $i$ corresponds to a crossed element that is not in the ICC, it will experience 
the action of $\hat{\mathbb{R}}_i(\theta/2,-\bm{n}_\perp) \hat{\mathbb{R}}_i(\theta/2,\bm{n}_\perp) =
\hat{\openone}_i$; on the contrary if $j$ happens to be the crossed element of the ICC it will undergo the transformation
\begin{eqnarray} 
 \hat{\sigma}_j^{(z)} \hat{\mathbb{R}}_{j}(\theta/2,-\bm{n}_\perp) \hat{\sigma}_j^{(z)} \hat{\mathbb{R}}_j(\theta/2,\bm{n}_\perp) =
  \hat{\mathbb{R}}_{i}(\theta/2,\bm{n}_\perp) \hat{\mathbb{R}}_j(\theta/2,\bm{n}_\perp)= 
  \hat{\mathbb{R}}_j(\theta,\bm{n}_\perp)\;. 
 \end{eqnarray} 

\subsection{Two-qubit gate}
As mentioned in the main text, to entangle two logical qubits we simply need to bring the interface at the position of the $A$-type crossed-qubit that connects the two logical qubits we want to entangle. Subsequently, we send a global pulse on the $A$-type qubits, designed to perform a $2\pi$ rotation on the crossed $A$-type qubits, i.e. 
the gate $\hat{Z}_{A^{\times}}^{(\rm tot)}$  defined in Eq.~(\ref{defZTOT}). 
 Thus, such qubits will acquire a $(-1)$ phase factor if and only if the two connected $B$- or $C$-type qubits are in the ground state, realizing a controlled-phase gate (CZ). Actually, the pulse is realized to perform a $2\pi$ rotation also on the normal $B$-type qubits. However, this simply accounts for a global phase factor, which we discard. In summary, the transformation 
 $\hat{Z}_{A^{\times}}^{(\rm tot)}$ can be realized through a five step sequence $\hat{U}^{(5)}_{A} \hat{U}^{(4)}_{A}\hat{U}^{(3)}_{A} \hat{U}^{(2)}_{A}\hat{U}^{(1)}_{A}$
 where 
\begin{itemize}
    \item $\hat{U}^{(1)}_{A}$ is induced by setting $\phi_A^{(1)} = \pi/2$ and letting evolving the system 
    for a time $\tau_1= \frac{\pi}{4} \Omega_A^{-1}$;
    \item $\hat{U}^{(2)}_{A}$ is induced by setting $\phi_A^{(2)} = 0$ and letting evolving the system 
    for a time $\tau_2 = \pi \Omega_A^{-1}$; 
    \item $\hat{U}^{(3)}_{A}$ is induced by setting $\phi_A^{(3)} = \pi/2$ and letting evolving the system 
    for a time $\tau_3= \frac{\pi}{2} \Omega_A^{-1}$;
    \item $\hat{U}^{(4)}_{A}$ is induced by setting $\phi_A^{(4)} = 0$ and letting evolving the system 
    for a time $\tau_4 = \pi \Omega_A^{-1}$;
        \item $\hat{U}^{(5)}_{A}$ is induced by setting $\phi_A^{(5)} = \pi/2$ and letting evolving the system 
    for a time $\tau_5 =  \frac{\pi}{4} \Omega_A^{-1}$.
\end{itemize}

\subsection{Initialization}\label{sec:iniz}
\begin{figure}[t!]
\centering
\includegraphics[width=0.8\columnwidth]{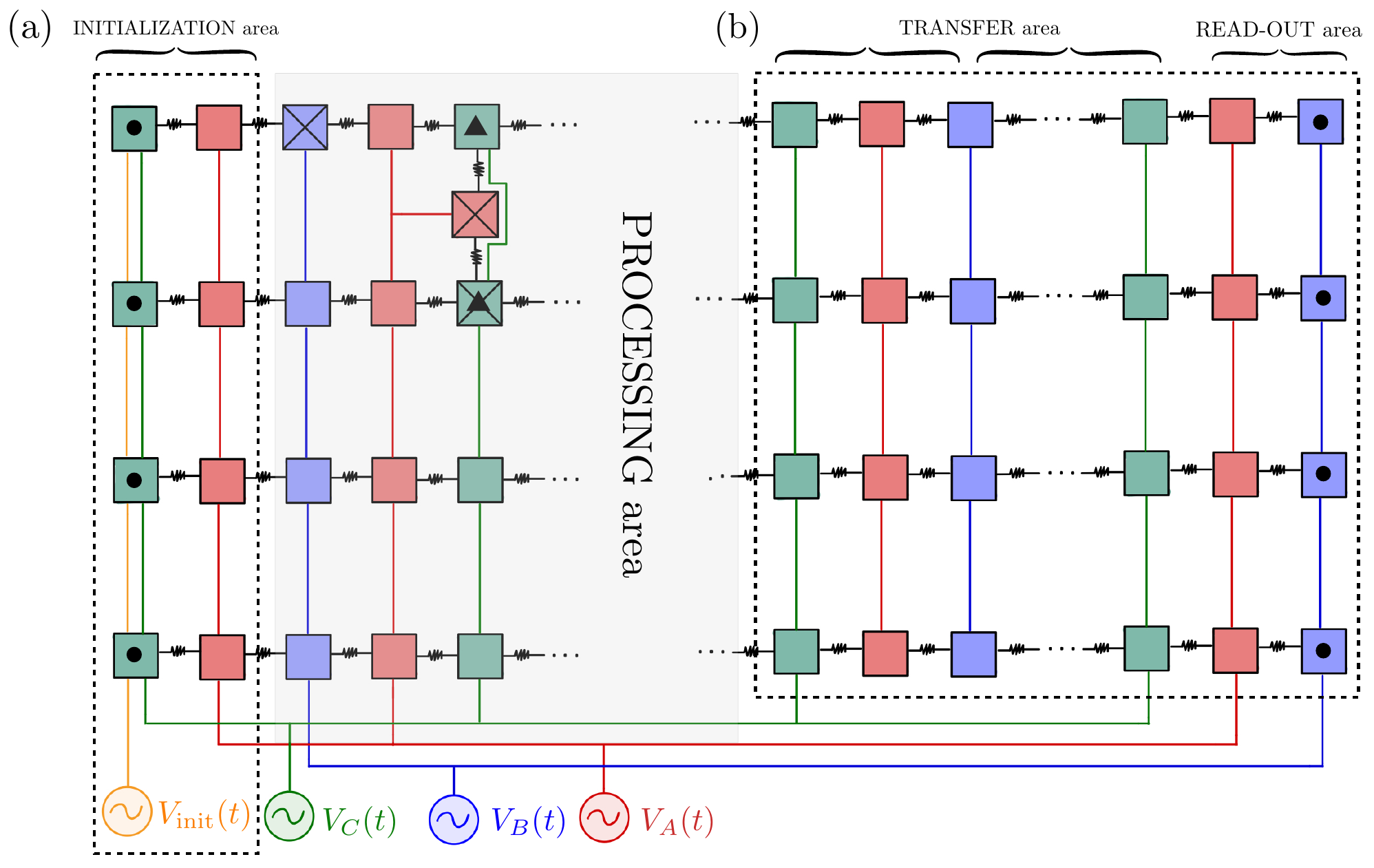}
\caption{In panel (a) we present an alternative realization of the initialization area with respect to the design of Fig.~1 of the main text. Here the first column is selectively addressed by an extra control line $V_{\rm init}(t)$ (continuous golden line) which acts independently from $V_A(t)$ and $V_B(t)$. Notice that in this case we do not need  the element of the first column to be crossed. In panel (b) we show how to use extra columns (transfer area) to outdistance the last column of the read-out area from the processing area.
\label{fig:alternativeini}}
\end{figure}

Before we start the computation,  the device is in a full ferromagnetic 
phase 
\begin{eqnarray} |\Psi_{\rm ferro}\rangle:= |g^{\otimes N}\rangle\otimes |g^{\otimes N}\rangle \cdots\;, \end{eqnarray}
 where all the qubits  are in the $|g\rangle$ state. This is a stable configuration
of the system Hamiltonian and, in case the $\zeta$ interaction term is positive, also corresponds to the ground state of the model. 
However, this is not a well-formed state and it will not react well when we apply the sequences~(\ref{stringSUP}).
An essential ingredient of our architecture is the ability to force transitions
from $|\Psi_{\rm ferro}\rangle$ to one of the vectors $|\Psi; {k}\rangle$. 
For this purpose we can use the following sequence of operations:
\begin{eqnarray} 
\hat{U}_{\rm init} : = \hat{\Pi}_{A^{\times}} \hat{\Pi}_{C^{\times}}\hat{\Pi}_{A^{\times}}\;,
\end{eqnarray} 
with $\hat{\Pi}_{A^{\times}}$ and $\hat{\Pi}_{C^{\times}}$ the controlled $\pi$-gates defined in Eq.~(\ref{dfdsfadf}). 
We notice that the action of the first $\hat{\Pi}_{A^{\times}}$ will promote (up to a global phase) the state 
$|\Psi_{\rm ferro}\rangle$ in a vector  $|\Psi'_{\rm ferro}\rangle$  where all the crossed $A$-type qubits are in the $|e\rangle$ state keeping all the others
in the $|g\rangle$ state. When acting on such configuration with $\hat{\Pi}_{C^{\times}}$ we will induce a $|g\rangle\rightarrow |e\rangle$ transition on the elements of the first column of the device since they are crossed $C$-type qubits with all neighbouring qubits in the ground state. Notice that the other crossed $C$-type qubits have instead a crossed $A$-type qubit in the $|e\rangle$ state
which blocks such transition. The new state $|\Psi''_{\rm ferro}\rangle$ has hence all the qubits of the first $C$ column
and all the crossed $A$-type elements in the $|e\rangle$ state and all the remaining elements in the $|g\rangle$ state. When we finally act with the
second  $\hat{\Pi}_{A^{\times}}$ operation, all the crossed elements of $A$ will return to $|g\rangle$ since all their neighbouring qubits are still in the $|g\rangle$ state, i.e. 
\begin{eqnarray} 
 |\Psi_{\rm ferro}\rangle &\stackrel{\hat{\Pi}_{A^{\times}}
 }{\longrightarrow}& |\Psi'_{\rm ferro}\rangle := \Big(   \otimes_{i\in A^{\times}} |e_i\rangle\Big) \Big(\otimes_{i'\notin A^{\times}}|g_{i'}\rangle \Big)\nonumber \\
&\stackrel{
\hat{\Pi}_{C^{\times}} }{\longrightarrow}&|\Psi''_{\rm ferro}\rangle:=  \Big(\otimes_{j\in {\cal C}_1}|e_j\rangle \Big)
\Big(   \otimes_{i\in A^{\times}} |e_i\rangle\Big) \Big(\otimes_{i'\notin A^{\times}\cup  {\cal C}_1}|g_{i'}\rangle \Big)\nonumber \\
&\stackrel{
\hat{\Pi}_{A^{\times}} }{\longrightarrow}&|\Psi'''_{\rm ferro}\rangle :=  \Big(\otimes_{j\in {\cal C}_1}|e_j\rangle \Big)
 \Big(\otimes_{i'\notin  {\cal C}_1}|g_{i'}\rangle \Big)\;,\label{ini0} 
 \end{eqnarray} 
 where ${\cal C}_1$ represents the set of qubits of the first column of the device. 
Notice that the final state of the transformation corresponds to the well-formed state $|\Psi_0; {3}\rangle$ where the ICC is located in the first $B$ column
of the processing area, and a logical state $|\Psi_0\rangle:= |g^{\otimes N}\rangle$. From this state, we can start the computation moving the ICC back and forth
on the processing area to apply any desired quantum gate.
\\
It is worth mentioning that an alternative approach to realize the mapping from  $|\Psi_{\rm ferro}\rangle$ to $|\Psi_0,3\rangle$, is to include an extra dedicated control line $V_{\rm init}(t)$   that acts selectively on the first column of the device as shown in panel (a) of Fig.~\ref{fig:alternativeini}. Such term contributes to the Hamiltonian of the model via the following
time-dependent term
 \begin{eqnarray}
\hat{H}_{\rm init}(t)&:=&  \sum_{i \in{\cal C}_1} \frac{\hbar \Omega_{\rm init}(t)}{2}  \Big(e^{i\phi_{\rm init}(t)}\vert g_i \rangle \langle e_i \vert + {\rm H.c.} \Big) \;, 
\label{HfinalSUPint} 
\end{eqnarray}
where  $\phi_{\rm init}(t)$ and 
$\Omega_{\rm init}(t)$ are the associated control functions.
Therefore  tailoring the phase, the Rabi frequency, and the duration $\tau_{\rm init}$ of the control line we can induce the evolution 
$\hat{\Pi}_{{\cal C}_1}:= \hat{W}_{{\cal C}_1}( \pi ; \bm{x})=
 \prod_{i\in {\cal C}_1}   \left[  \hat{\openone}_i \otimes \hat{Q}_{\langle i \rangle}- i \hat{\sigma}^{(x)}_i\otimes    \hat{P}_{\langle i \rangle} \right]$, 
which indeed allow us to realize the required mapping. 

\subsection{Read-out}\label{sec:redout}
At the end of the computation, we read out the logical state of the ICC by moving it to the rightmost element of the process unit (i.e., the $B$- or $C$-type column of the read-out area) using the transformations $\hat{U}_{\rm shift}$. Despite it is theoretically possible to measure the logical state of the ICC column while it is inside the process area, we prefer not to do this. Such an approach would inevitably require local addressing of the elements within the column, which contradicts the fundamental principles of our architecture.
Notice also that the scheme presented in Fig.~1 of the main text represents the minimal setting that allows for a separation between the processing area and the read-out column. If we need to increase this separation, it can be easily accomplished by adding an extra sequence of columns in the read-out area, see panel (b) of Fig.~\ref{fig:alternativeini}.

\section{Numerical Simulations} \label{sec:Numerical Simulations}
In this Section we perform exact numerical simulations to test the $e-e$ blockade regime (Sec.~\ref{sec:simu-blockade}) and the validity of the protocol allowing the motion of logical states (Sec.~\ref{sec:sim-motion}) and the protocol realizing the Hadamard (single-qubit) gate (Sec.~\ref{sec:sim-hadamard}).

\subsection{Testing the blockade regime}\label{sec:simu-blockade}
\begin{figure}[!t]
\centering
\includegraphics[scale=0.75]{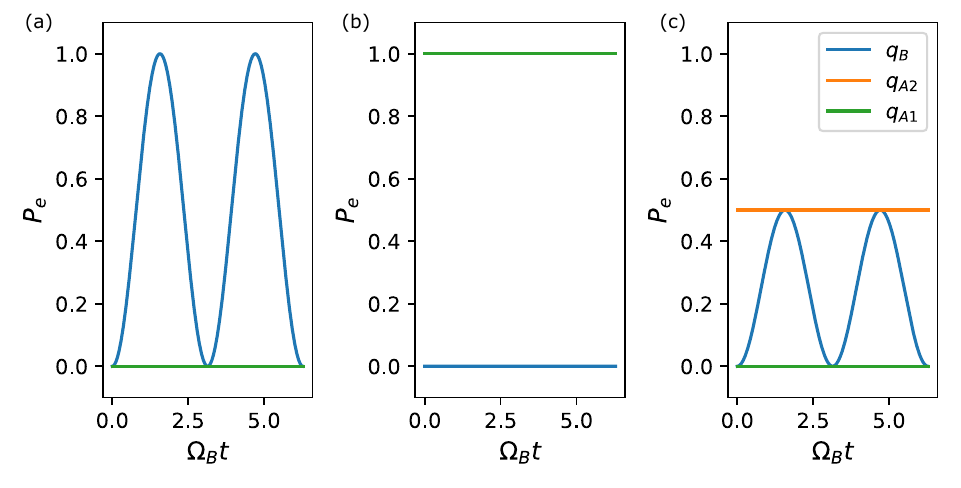}
\caption{\label{fig:sim-ee_drive} Numerical simulation for a three-qubit chain where the $B$-type qubit is the only driven by an external $\pi$-pulse. Here $P_e$ denotes the population of the excited state of each qubit and time is measured in units of $\Omega_B$. The driven qubit is denoted as $q_B$ and is initialised in the ground state. Its neighbours are $q_{A1}$ and $q_{A2}$. To reach the $e$--$e$ blockade, pulses of frequency $\omega_{\rm d} =\omega_B - 2 \zeta$ drive the $B$-type qubit of the chain. The plots in the figure show the population $P_e$ after the effect of the driving pulse. (a) The qubits at the extremes of the chain are initialised in the ground state, so the dynamics of the driven qubit is not blocked. (b) The external qubits are both in the excited state, resulting in blockade dynamics for the driven $B$-type qubit. (c) The qubit on the right of the chain, $q_{A2}$, is initialised in the superposition $\vert + \rangle$, while the qubit $q_{A_1}$ is in the ground state. With this initial configuration, we expect the driven qubit to reach the $\vert + \rangle$ state, since it is “half" blocked. }
\end{figure}

In order to test our proposal for reaching the blockade effect, we simulate the dynamic evolution of a system made of three qubits, e.g. the chain $A_1BA_2$, under the Hamiltonian~\eqref{Htotsup}. Here, the drive~\eqref{HdriveSUP} controls only the $B$-type qubit in the middle of the chain, whose natural frequency is $\omega_B$. In Fig.~\ref{fig:sim-ee_drive} the dynamics of the $B$-type qubit is shown. As explained (e.g.~see Fig.~2(a) of the main text), in order to reach the $e$--$e$ blockade, we must drive the $B$-type qubit with an off-resonant pulse of frequency $\omega_{\rm d} =\omega_B - 2 \zeta$. In the following, we simulate the system starting from three different initial configurations: $\vert ggg\rangle$, $\vert gge\rangle$ and $\vert gg +\rangle$~\cite{+state}. The plots of Fig.~\ref{fig:sim-ee_drive}, from left to right, show the population $P_e$ of the excited state over the $\pi$-pulse dynamics for a three-qubit chain. It is easy to see that the desired blockade regime is reached.

\subsection{Checking the motion of the interface}\label{sec:sim-motion}
\begin{figure}[!t]
\centering
\includegraphics[scale=0.5]{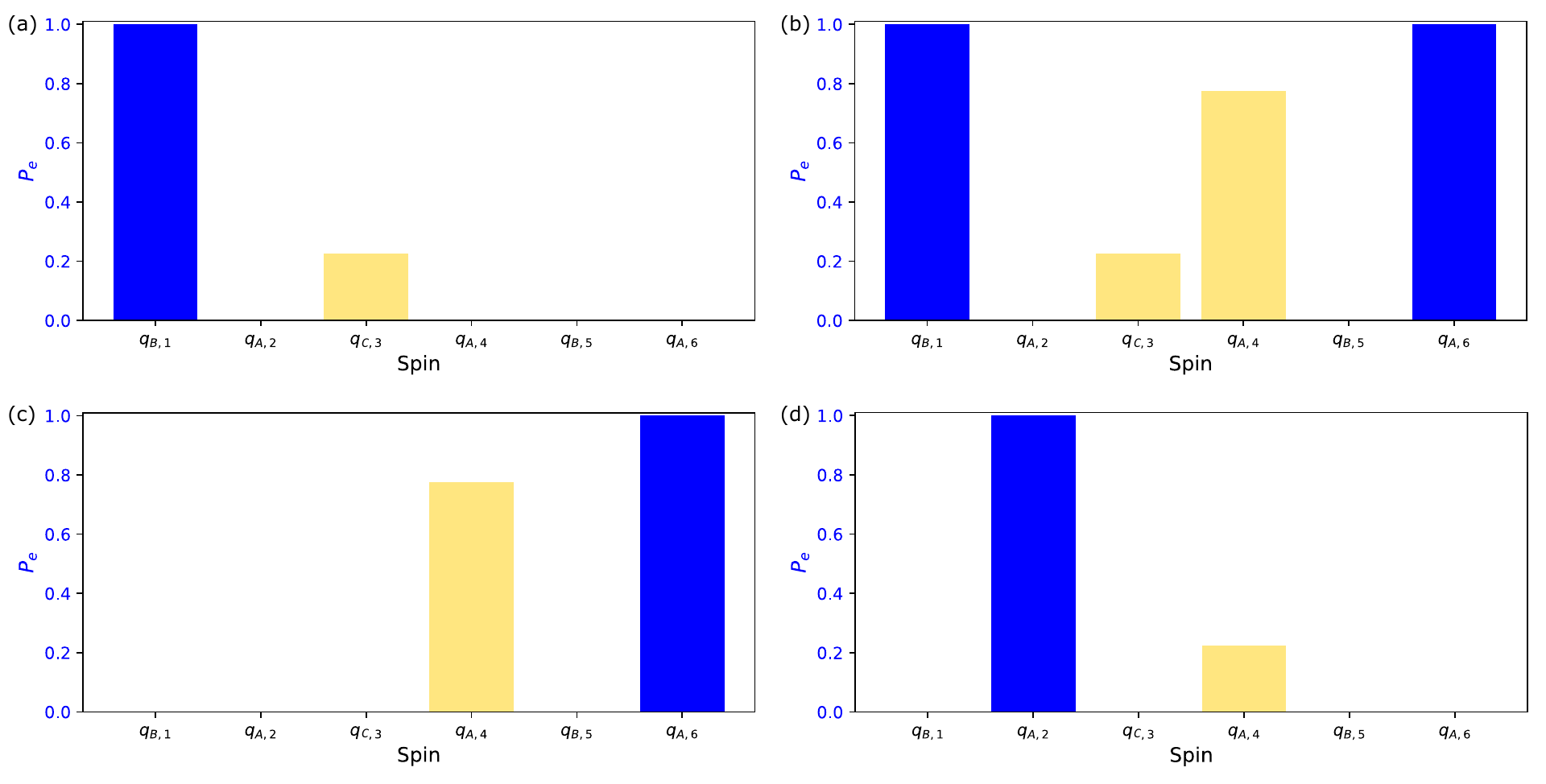}
\caption{\label{fig:sim-ee} Time evolution for a single row of 6 regular qubits. The bars indicate the probability $P_e$ of finding the two-level system in the excited state. The $\chi$-type qubits in the row are denoted as $q_{\chi,i}$, where $i\in(1,6)$ stands for the spatial position coordinate in the row. (a) Initial state of the system, with the $i=3$ qubit as the logic one. The interface is colored in yellow. (b) State of the system after the $\hat{\Pi}_A$ pulse. Here the third and fourth qubits are entagled and the interface is shared. (c) State of the system after the $\hat{\Pi}_B\hat{\Pi}_C\hat{\Pi}_A$ pulse. (c) Final state, after the entire pulse sequence $\hat{U}_{\rm shift}= \hat{\Pi}_A\hat{\Pi}_B\hat{\Pi}_C\hat{\Pi}_A$. The logical qubit is now at position $i=4$.}
\end{figure}

By using the protocol explained in the previous Section, we now perform numerical simulations in order to check the correct motion of the interface for one single row of alternating $A, B$- and $C$- type qubits. There are no crossed qubits in the system. Firstly, we initialize the state of the qubits and then we allow the system to  evolve under the total Hamiltonian~\eqref{HfinalSUP}. Secondly, we perform the protocol defined above in Sec.~\ref{sec:motioninter}. The results are shown in Fig.~\ref{fig:sim-ee} where the interface is shifted from left to right.

\subsection{Gate Fidelity}
\label{sec:sim-hadamard}
\begin{figure}[!t]
\centering
\includegraphics[scale=0.8]{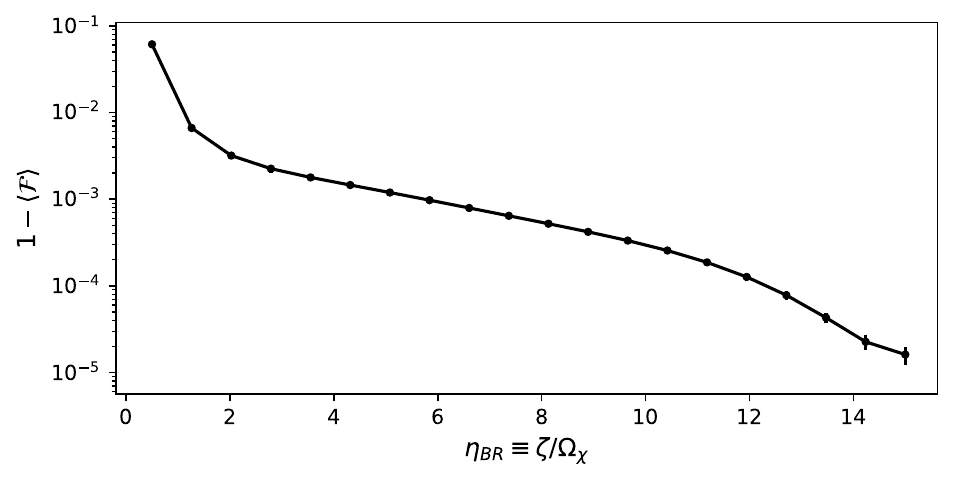}
\caption{\label{fig:fidelity} $1- \langle\mathcal{F}\rangle$ as a function of the ratio between the ZZ coupling $\zeta$ and the Rabi frequency $\Omega_\chi$. Note that $\eta_{\rm BR}=5$ is sufficient to reach a $99.9\%$ gate fidelity.}
\end{figure}

One method for evaluating the efficacy of our protocol is to compute the average gate fidelity, denoted by $\langle\mathcal{F}\rangle$. It is possible that errors are introduced due to the blockade being exact only in the $\eta_{\rm BR} \gg 1$ limit. In particular, using a 5-qubits row, we simulate the pulse sequence which performs a Hadamard gate on the logical qubit at the position of the crossed $B$-type qubit, see Sec.~\ref{sec:single-qubit}. The average is calculated by averaging over approximately $\mathcal{O}(10^2)$ random initializations of the crossed qubit state. For each initialization, the fidelity is computed according to the formula~\cite{Nielsen2010}
\begin{equation}
\mathcal{F}(\hat{\rho}_{\mathrm{target}}, \hat{\rho}_{\mathrm{exp}}) = \mathrm{Tr}\left(\sqrt{\sqrt{\hat{\rho}_{\mathrm{target}}}\hat{\rho}_{\mathrm{exp}}\sqrt{\hat{\rho}_{\mathrm{target}}}}\right) \ ,
\end{equation}
where the density matrix of the system after the application of a perfect Hadamard gate on the crossed qubit is denoted by $\hat{\rho}_{\mathrm{target}}$, while the density matrix we get after performing the pulse sequence in our system is represented by $\hat{\rho}_{\mathrm{exp}}$. The result is presented in Fig.~\ref{fig:fidelity}, in which the average gate fidelity is plotted as a function of the ratio between the qubit-qubit coupling constant, $\zeta$, and the Rabi frequency, $\Omega_\chi$, with $\chi = A,B,C$. From the graph, it is evident that a ratio of 5 is sufficient to achieve a gate fidelity of $99.9\%$.
